\documentclass[a4paper,11pt,fleqn]{article}

\usepackage[ansinew]{inputenc}
\usepackage{hyperref}
\usepackage[mathscr]{eucal}
\usepackage{amsmath,amssymb,amsthm}
\usepackage{graphicx}
\usepackage{color}

\allowdisplaybreaks

\hypersetup{colorlinks, linkcolor=blue, citecolor=blue, urlcolor=blue}
\newcommand{\ddd}{\mathrm{d}}
\newcommand{\dddh}{\mathrm{d}_{\mathrm h}}
\newcommand{\p}{\partial}
\newcommand{\const}{\mathop{\rm const}\nolimits}
\newcommand{\sign}{\mathop{\rm sgn}\nolimits}

\newcommand{\todo}[1][\null]{\ensuremath{\clubsuit}}

\newcommand{\noprint}[1]{}

\newtheorem{theorem}{Theorem}

{\theoremstyle{definition}

\newtheorem{example}{Example}
\newtheorem{remark}{Remark}
\newtheorem*{remark*}{Remark}
}

\newcommand{\checked}[1][\null]{\ensuremath{\boldsymbol{\surd}}}

\newcommand{\ff}{\mathsf{f}}
\newcommand{\DD}{\mathrm{D}}

\newcommand{\vv}{\mathbf{v}}

\newcommand{\ve}{\varepsilon}

\newcommand{\nn}{\nabla}

\newcommand{\ZZ}{\mathcal{Z}}

\newcommand{\XX}{\mathcal{X}}
\newcommand{\DDD}{\mathcal{D}}

\begin{document}

\par\noindent {\LARGE\bf
Invariant parameterization and turbulence modeling\\ on the beta-plane
\par}

{\vspace{6mm}\par\noindent {\large Alexander Bihlo$^{\dag1}$, Elsa Dos Santos Cardoso-Bihlo$^{\ddag2}$ and Roman O.\ Popovych$^{\ddag\S3}$
} \par}

{\vspace{6mm}\par\noindent {\it
$^{\dag}$~Centre de recherches math\'{e}matiques, Universit\'{e} de Montr\'{e}al, C.P.\ 6128, succ.\ Centre-ville,\\
$\phantom{^\dag}$~Montr\'{e}al (QC) H3C 3J7, Canada
}}

{\vspace{2mm}\par\noindent {\it
$^{\ddag}$~Wolfgang Pauli Institute, Oskar-Morgenstern-Platz 1, A-1090 Vienna, Austria
}}

{\vspace{2mm}\par\noindent {\it
$^{\S}$~Institute of Mathematics of NAS of Ukraine, 3 Tereshchenkivska Str., 01601 Kyiv, Ukraine
}}

{\vspace{2mm}\par\noindent{\textup{E-mail}:
$^1$bihlo@crm.umontreal.ca,
$^2$elsa.cardoso@univie.ac.at,
$^3$rop@imath.kiev.ua
}\par}

\vspace{8mm}\par\noindent\hspace*{5mm}\parbox{150mm}{\small
Invariant parameterization schemes for the eddy-vorticity flux in the barotropic vorticity equation on the beta-plane are constructed and then applied to turbulence modeling.
This construction is realized by the exhaustive description of differential invariants for the maximal Lie invariance pseudogroup of this equation using the method of moving frames,
which includes finding functional bases of differential invariants of arbitrary order, a minimal generating set of differential invariants and a basis of operators of invariant differentiation in an explicit form.
Special attention is paid to the problem of two-dimensional turbulence on the beta-plane.
It is shown that classical hyperdiffusion as used to initiate the energy--enstrophy cascades violates the symmetries of the vorticity equation. Invariant but nonlinear hyperdiffusion-like terms of new types are introduced and then used in the course of numerically integrating the vorticity equation and carrying out freely decaying turbulence tests.
It is found that the invariant hyperdiffusion scheme is closely reproducing the theoretically predicted $k^{-1}$ shape of enstrophy spectrum in the enstrophy inertial range.
By presenting conservative invariant hyperdiffusion terms, we also demonstrate that the concepts of invariant and conservative parameterizations are consistent.
}\par\vspace{5mm}

\noprint{
PACs:
47.10.A- Mathematical formulations
47.10.ab Conservation laws and constitutive relations
47.27.E- Turbulence simulation and modeling
47.27.em Eddy-viscosity closures; Reynolds stress modeling
92.60.hk Convection, turbulence, and diffusion (see also 92.30.Ef�in Geophysics Appendix)

}

\section{Introduction}

As atmospheric and oceanic numerical models get increasingly complex, it becomes more and more challenging to propose valuable conceptual paradigms for those processes that the model is still not able to capture owing to its limited spatial and temporal resolution. This problem is common to all numerical models irrespective of their eventual degree of sophistication~\cite{sten07Ay,stul88Ay}. In the beginning of numerical modeling in geophysical fluid dynamics, it was often the lack of computer power that dictated which processes had to be parameterized, even with a concise understanding of these processes. As computers became more capable, the problem of parameterization shifted to processes occurring on rather fine scales where it can be difficult to retrieve accurate experimental data. Accordingly, for various processes that should be taken into account in order to improve the forecast range of a numerical model, there is still no satisfactory general understanding. This naturally makes it difficult to set up valuable parameterization schemes, which for this reason is usually an elaborate~task.

\looseness=1
On the other hand, processes that occur in geophysical fluid dynamics and that can be described using differential equations also might have certain structural or geometrical properties. Such properties can be conservation of mass or energy or other fundamental conservation laws. Real-world processes are generally also invariant under specific transformation groups, as e.g.\ the Galilean group. This is why one can ask the question whether it is reasonable to construct parameterization schemes for processes possessing certain structural features in a manner such that these features are preserved in the closed model. In this way, even if a model is not able to explicitly resolve processes, loosely speaking, it takes into account some of their significant properties. This study was initiated in~\cite{ober97Ay} for the problem of finding invariant turbulence closure schemes for the filtered Navier--Stokes equations. In the present paper we aim to give a further instance for invariant parameterization schemes by constructing closure ansatzes that retain certain Lie point symmetries of the barotropic vorticity equation on the beta-plane.

This possible stream of constructing geometrically motivated parameterization schemes in some sense parallels the present general trend in numerical modeling to design specially adapted discretizations of differential equations that capture a range of their qualitative or global features, such as conservation laws, a Hamiltonian structure or symmetry properties. Especially the possibility of constructing discretization schemes that have the same symmetries and/or conservation laws as the original differential equations they are a model of, as proposed and discussed e.g.\ in~\cite{bihl12Ay,doro11Ay,kim06Ay,kim08Ay,rebe11Ay,salm05Ay,salm07Ay,vali05Ay}, is of immediate relevance to the present work. This is because, strictly speaking, a discretization of a system of differential equation is in practice not enough to set up a valuable numerical model. There always has to be a model for the unresolved parts of the dynamics. (Neglecting them is in general not a good idea as for nonlinear differential equations these parts will, sooner or later, spoil the numerical integration.) Then, if one aims to construct an invariant discrete counterpart of some relevant physical model, care should also be taken about the invariance characteristics of the processes that are not explicitly resolved. This is where the method of finding invariant parameterization schemes comes into play. The combination of invariant discretization schemes for the resolved part of the model with invariant parameterization schemes for the unresolved parts yields a completely invariant numerical description of a given system of differential equations. Such a fully invariant model might be closer to a true geometric numerical integration scheme than solely a symmetry preserving discretization without any closure for the subgrid-scale terms or with some non-invariant closure.

Perhaps the most relevant usage of the barotropic vorticity equation is related to two-dimensional turbulence. Turbulence on the beta-plane (or, more general, on the rotating sphere) is peculiar in that it allows for the combination of turbulent and wave-like effects. It is believed to explain the emergence of strong jets and band-like structure on giant planets in our solar system and is therefore the subject of intensive investigation, see e.g.~\cite{huan01Ay,malt91Ay,rhin75Ay,salm98Ay,vall06Ay} and references therein. In the present paper we focus on freely decaying turbulence on the beta-plane by using invariant hyperdiffusion terms to initiate the energy--enstrophy cascades. These cascades are likely responsible for the emergence of coherent, stable structures (vortices) that are ubiquitous in large-scale geophysical fluid dynamics. 
Note that the possibility of a development of coherent structures causing the classical inverse cascade to break down is also discussed in the literature.  Energy can then be transferred between different scales without a non-linear cascade~\cite{robi06Ay}. 

\looseness=1
The outline of the paper is as follows. In the subsequent Section~\ref{sec:Theory}, we discuss and slightly extend the concept of invariant parameterization schemes as introduced in~\cite{ober97Ay} and~\cite{popo10Cy}. Special attention will be paid to methods related to invariant parameterization schemes and inverse group classification.
In Section~\ref{sec:SymmetriesVorticityEquation} we present the maximal Lie invariance algebra~$\mathfrak g_1$ and the maximal Lie invariance pseudogroup of the barotropic vorticity equation on the beta-plane. The computation of the algebra~$\mathfrak g_1$ is briefly described in Appendix~\ref{sec:AppendixSymmetriesVorticityEquation}.
A concise description of the general method for computing differential invariants of Lie (pseudo)groups using the method of moving frames is given in Section~\ref{sec:AlgorithmOfConstructionOfDiffInvs}. In Section~\ref{sec:DifferentialInvariantsOfVorticityEquation} the algebra of differential invariants is determined for the maximal Lie invariance pseudogroup of the vorticity equation. The related computation can be found in Appendix~\ref{sec:AppendixStructureAlgebraDifferentialInvariants}.
Two examples for invariant parameterization schemes constructed out of existing schemes using the invariantization process are presented in Section~\ref{sec:InvariantizationOfBetaPlaneBVE}.
Section~\ref{sec:ApplicationOfInvariantParameterization} is devoted to the application of differential invariants in turbulence on the beta-plane. In particular, invariant hyperdiffusion schemes are introduced. The vorticity equation on the beta-plane is integrated numerically using both invariant and non-invariant hyperdiffusion and the corresponding enstrophy spectra are obtained.
In Section~\ref{SectionOnConservativeInvariantParameterization} we discuss the possibility of deriving invariant parameterization schemes that also respect conservation laws. As an example, an invariant diffusion term is constructed that preserves the entire maximal Lie invariance pseudogroup of the vorticity equation and also preserves conservation of energy, circulation and momentum.
The results are summarized and further discussed in Section~\ref{sec:Conclusion}, in which we also indicate possible future research directions in the field of invariant parameterization.

\section{Invariant parameterization schemes}\label{sec:Theory}

The problem of finding parameterization or closure schemes for subgrid-scale terms in averaged differential equations that admit Lie symmetries of the original (unaveraged) differential equation was first raised in~\cite{ober97Ay}, see also~\cite{ober00Ay,raza07Ay}. Recently, we put this idea into the framework of group classification~\cite{popo10Cy}, by showing that any problem of constructing invariant parameterization schemes amounts in solving a (possibly complicated) group classification problem.

As for the classical group classification, there are two principal ways to construct parameterization schemes, the direct and the inverse one~\cite{popo10Cy}. In the direct approach, one replaces the terms to be parameterized with arbitrary functions depending on the mean variables and derivatives thereof. This is in the line with the general definition of all physical parameterization schemes, which are concerned to express the unknown subgrid-scale terms using the information included in the grid-scale (mean) quantities. The form of dependency of the arbitrary functions on the mean variables is guided by physical intuition. It determines the properties of all the families of invariant parameterization schemes that can be derived (e.g.\ the highest order of derivatives that can arise). Once the general form of the arbitrary function is chosen, one is left with a possibly rather general class of differential equations, which is amenable with tools from usual group classification, see e.g.~\cite{bihl11Dy,card11Ay,popo10Ay}. This in particular will lead to a list of families of mutually inequivalent parameterization schemes that admit different Lie invariance algebras. One can then select those families that preserve the most essential symmetry features of the process to be parameterized. The final step is to suitably narrow the selected families by including other desirable physical properties into the invariant parameterization scheme.

In the present paper, however, we will be solely concerned with the inverse approach, which is why we will discuss it in a more extended manner. The inverse approach rests on the fact that any system of differential equations can be rewritten in terms of differential invariants of its maximal Lie invariance group, provided that the prolongation of the group to the corresponding jet space acts semi-regularly~\cite{olve86Ay}. This property can be used in the course of the parameterization problem in the following way: Suppose that we are given a Lie group~$G$ regarded as important to be preserved for valuable parameterization schemes as a Lie symmetry group. Computing a basis of differential invariants of~$G$ along with a complete set of its independent operators of invariant differentiation, see e.g.~\cite{fels98Ay,fels99Ay,olve07Ay,ovsi82Ay}, serves to exhaustively describe the entire algebra of differential invariants of~$G$. As any combination of these differential invariants will necessarily be invariant with respect to~$G$, assembling them together to a parameterization will immediately lead to a closure scheme admitting~$G$ as a Lie symmetry group.

The key question hence lies in the correct selection of a suitable symmetry group.
The initial point for the selection is given by symmetry properties of the model to be parameterized.
In the course of the parameterization one can intend to preserve the whole Lie symmetry group of the initial model or its proper subgroups.
The choice for an invariance group for parameterization obviously should not solely be justified using mathematical arguments. Sometimes, it can be motivated from obvious physical reasons. If the process to be parameterized can be described within the framework of classical mechanics then any reasonable parameterization for that process should be invariant under the Galilean group. Moreover, for turbulence closure schemes, scale invariance might be of particular importance.
For processes that can be described within the framework of a variational principle and respect certain conservation laws, it might be reasonable that the parameterization scheme to be developed respects the associated Noether symmetries. See~\cite[Chapter~6]{doro11Ay} for similar studies of discretization schemes within the variational framework.

There are several processes in fluid mechanics that are intimately linked to the presence of certain boundary conditions (e.g., turbulence near walls, boundary layer convection, etc.).
For such processes the inclusion of the particular boundaries is an integral part in the formulation of a parameterization scheme.
At first glance, to find invariant parameterization schemes for such processes it is inevitable
to single out those subgroups of the maximal Lie invariance group~$G$ of the system~$\mathcal L$ of differential equations describing the process of interest
that are compatible with a particular boundary value problem.
The main complication with this approach is that most of boundary value problems admit no symmetries, see e.g.~\cite{blum89Ay}.
At the same time, it is more natural to assume that symmetries of~$\mathcal L$ act as equivalence transformations on
a joint class of physically relevant boundary value problems for~$\mathcal L$,
i.e., these transformations send a particular boundary value problem to another problem from the same class~\cite{bihl12By}.
Even the basic physical symmetries including shifts in space and time, rotations, scalings, Galilean boosts or Lorentz transformations,
which are related to fundamental properties of the space and the time (homogeneity, isotropy, similarity, Galilean or special relativity principle, respectively),
usually act on boundary value problems in much the same way as equivalence transformations.
This is why it is the generation of a group of well-defined equivalence transformations on a properly chosen class of boundary value problems
that can serve as a criterion for selecting a subgroup of~$G$ to be taken into account in the course of invariant parameterization~of~$\mathcal L$.

Employing techniques of inverse group classification does not automatically lead to ready-to-use parameterizations, but it gives a frame in which parameterizations can be defined without the violation of basic invariance properties. Examples of the violation have been reported in the recent literature. See, e.g.,~\cite{ober97Ay} for a discussion about the Smagorinsky model in the filtered Navier--Stokes equations violating scale invariance and~\cite{popo10Cy} for a note on the Kuo convection schemes that describes a Galilean invariant process in a non-invariant fashion. The construction of parameterization schemes that fail to preserve essential symmetries can be easily avoided by applying the above methods of inverse group classification. This may help to restrict the large number of possible closure schemes using geometrical reasoning and thereby may assist in finding a proper description of unresolved subgrid-scale processes.

There is yet a second possibility to construct invariant parameterization schemes using the inverse group classification approach, which has not been reported in~\cite{popo10Cy}. It rests on the construction of \emph{moving frames} for the Lie group~$G$ with respect to which parameterizations under study should be invariant. It is a general feature of a moving frame that it allows constructing of invariant counterparts of differential functions. This property enables the construction of an invariant parameterization scheme out of a non-invariant one. It is simply necessary to apply the moving frame corresponding to the selected Lie group~$G$ to the specific closed differential equation. More precisely, consider a system~$\mathcal L$ of differential equations
\[
 L^l(x,u_{(n)})=0,\quad l=1,\dots,m.
\]
The dependent variables $u$ can be represented according to $u=\bar u+u'$, where $\bar u$ is the averaged or filtered part of the dynamics (i.e.\ the resolved or grid-scale part) and $u'$ denotes the departure of $u$ from the mean or filtered part $\bar u$ (i.e.\ the subgrid-scale part). Numerical models in geophysical fluid dynamics are formulated as equations for the resolved part, which are obtained from $L^l=0$ by averaging or filtering, leading to
\[
 \tilde L^l(x,\bar u_{(n)},w)=0,\quad l=1,\dots,m,
\]
where $\tilde L^l$ are smooth differential functions of their arguments. The particular form of $\tilde L^l$ depends on the actual averaging rule chosen and the form of the initial system~$\mathcal L$. The unknown subgrid-scale terms that arise in the course of averaging (e.g.\ by using the Reynolds averaging rule for products, $\overline{ab}=\bar a\bar b+\overline{a'b'}$) are collected in the tuple~$w$. These terms have to be parameterized in order to close the above system of averaged differential equations. A local parameterization scheme establishes a particular functional relation
\[
 w=\theta(x,\bar u_{(r)})
\]
between the subgrid-scale and grid-scale quantities. Let there be given a moving frame $\rho^{(j)}$ of order $j=\max(r,n)$ for the selected Lie group~$G$, see Section~\ref{sec:AlgorithmOfConstructionOfDiffInvs}. Any particular parameterization scheme can then be invariantized via replacing $\tilde L^l$ and $\theta$ by their invariantized counterparts,
\[
\iota(\tilde L^l(x,\bar u_{(n)},w))=\tilde L^l(\rho^{(j)}\cdot x,\rho^{(j)}\cdot\bar u_{(n)},w)\quad\mbox{and}\quad
\iota(\theta(x,\bar u_{(r)}))=\theta(\rho^{(j)}\cdot x,\rho^{(j)}\cdot\bar u_{(r)}).
\]

\begin{example}
 It is instructive to illustrate this idea with a simple example. Let us consider the famous Korteweg--de Vries (KdV) equation,
\[
u_t+uu_x+u_{xxx}=0.
\]
Its maximal Lie invariance group $G_{\rm KdV}$ is four-dimensional and the most general transformation leaving the KdV equation invariant is
\begin{equation}\label{eq:GeneralSymmetryTransformationKdV}
 (T,X,U)=(e^{3\ve_4}(t+\ve_1),e^{\ve_4}(x+\ve_2+\ve_1\ve_3+\ve_3t),e^{-2\ve_4}(u+\ve_3)),
\end{equation}
where $\ve_1$, \dots $\ve_4$ are arbitrary constants.
Let us now apply the classical Reynolds averaging to the KdV equation. This yields
\[
 \bar u_t+\bar u\bar u_x+\bar u_{xxx}=-\frac12(\overline{{u'}^2})_x,
\]
where the right-hand side is the term we seek closure for. A simple closure ansatz is the down-gradient parameterization, i.e.\ we close the above equation by setting $\overline{u'^2}/2=-\kappa \bar u_x$, where for the sake of simplicity we use $\kappa=\const$. This yields the closed KdV equation
\begin{equation}\label{eq:ClosedKdV}
\bar u_t+\bar u\bar u_x+\bar u_{xxx}=\kappa \bar u_{xx}.
\end{equation}
However, it is easily verified that this equation is not invariant under the transformation~\eqref{eq:GeneralSymmetryTransformationKdV}. Namely, the scale invariance is lost, i.e.\ the closed KdV equation is invariant only under the three-parameter group of transformations associated with the group parameters $\ve_1$, $\ve_2$ and $\ve_3$. To restore scale invariance, we can invariantize the closed KdV equation~\eqref{eq:ClosedKdV} using the moving frame associated with the group $G_{\rm KdV}$.

Moving frames for the group $G_{\rm KdV}$ were constructed in~\cite{cheh08Ay,card13Ay}.
It is convenient to invariantize Eq.~\eqref{eq:ClosedKdV} using the moving frame with
\[
\ve_1=-t,\quad \ve_2=-x,\quad \ve_3=-\bar u,\quad \ve_4=\smash{\frac13}\ln \bar u_x.
\]
This is done by firstly applying the transformations~\eqref{eq:GeneralSymmetryTransformationKdV} to~\eqref{eq:ClosedKdV} which yields
\[
 \bar u_t+\bar u\bar u_x+\bar u_{xxx}=\kappa e^{\ve_4}\bar u_{xx},
\]
showing explicitly that this equation fails to be scale invariant. The invariantization is completed upon substituting the moving frame for $\ve_4$ giving
\[
 \bar u_t+\bar u\bar u_x+\bar u_{xxx}=\kappa \sqrt[3]{\bar u_x}\bar u_{xx}.
\]
It is readily checked that this closed equation is invariant under the same symmetry group $G_{\rm KdV}$ as is the original KdV equation. The price for restoring scale invariance of the closed KdV equation invoking the simple down-gradient parameterization is that the closure scheme becomes nonlinear. We will observe the same effect when invariantizing linear hyperdiffusion models for the vorticity equation on the beta-plane, which will be shown in detail below.
\end{example}

\section{Lie symmetries of the vorticity equation}\label{sec:SymmetriesVorticityEquation}

The barotropic vorticity equation on the beta-plane is a simple but still genuine meteorological model. It has the form
\begin{equation}\label{eq:VorticityEquation}
 \zeta_t+\psi_x\zeta_y-\psi_y\zeta_x+\beta\psi_x=0, \quad\mbox{or}\quad \zeta^{\rm a}_t + J(\psi,\zeta^{\rm a})=0.
\end{equation}
Here $J(a,b):=a_xb_y-a_yb_x$,
$\psi=\psi(t,x,y)$ is the stream function,
$\zeta=\psi_{xx}+\psi_{yy}$ is the vorticity
and $\zeta^{\rm a}=\zeta+\ff=\zeta+\ff_0+\beta y$ is the absolute vorticity under the $\beta$-plane approximation $\ff=\ff_0+\beta y$ of the Coriolis parameter~$\ff$,
$\beta$ is a nonzero constant parameter (the differential rotation).
The constant~$\ff_0$ is dynamically inessential and can be neglected.

The maximal Lie invariance algebra~$\mathfrak g_1$ of Eq.~\eqref{eq:VorticityEquation} is spanned by the vector fields
\[
\DDD=t\p_t-x\p_x-y\p_y-3\psi\p_\psi,\quad \p_t, \quad \p_y, \quad \XX(\tilde f)=\tilde f(t)\p_x-\tilde f_t(t)y\p_\psi \quad \ZZ(\tilde g)=\tilde g(t)\p_\psi,
\]
where the parameters $\tilde f$ and $\tilde g$ run through the set of smooth functions of~$t$~\cite{bihl09Ay,ibra95Ay}. More details on how the above vector fields are obtained can be found in Appendix~\ref{sec:AppendixSymmetriesVorticityEquation}. The vorticity equation~\eqref{eq:VorticityEquation} also admits two independent discrete symmetries, which alternate signs of the pairs $(t,x)$ and $(y,\psi)$, see~\cite{bihl11Cy} for more details. Such discrete symmetries will not be taken into account in the course of construction of differential invariants. Any nonzero value of~$\beta$ can be gauged to one by a scaling transformation.

The one-parameter Lie (pseudo)groups generated by the above vector fields read
\begin{align*}
 &\Gamma_{\ve_1}\colon\quad (t,x,y,\psi)\mapsto(e^{\ve_1}t,e^{-\ve_1}x,e^{-\ve_1}y,e^{-3\ve_1}\psi)\\
 &\Gamma_{\ve_2}\colon\quad (t,x,y,\psi)\mapsto(t+\ve_2,x,y,\psi)\\
 &\Gamma_{\ve_3}\colon\quad (t,x,y,\psi)\mapsto(t,x,y+\ve_3,\psi)\\
 &\Gamma_{f}\colon\quad (t,x,y,\psi)\mapsto(t,x+f(t),y,\psi-f_t(t)y)\\
 &\Gamma_{g}\colon\quad (t,x,y,\psi)\mapsto(t,x,y,\psi+g(t)),
\end{align*}
where $\ve_i\in \mathbb{R}$ and $f(t):=\ve_4\tilde f(t)$ and $g(t):=\ve_5\tilde g(t)$.
Accordingly, the admitted Lie symmetries of the barotropic vorticity equation on the beta-plane are scalings, time translations, translations in $y$-direction, generalized Galilean boosts in the $x$-direction and gaugings of the stream function with smooth time-dependent summands.

We will compose transformations from these one-parameter Lie (pseudo)groups in the following way
$
 \Gamma=\Gamma_{\ve_1}\circ\Gamma_{\ve_2}\circ\Gamma_{\ve_3}\circ\Gamma_{f}\circ\Gamma_{g}
$
to a transformation~$\Gamma$ from the maximal Lie symmetry pseudogroup~$G_1$ of the vorticity equation~\eqref{eq:VorticityEquation}. Any transformation of~$G_1$ then has the form
\begin{equation}\label{eq:GeneralTransformationVorticityEquation}
 (T,X,Y,\Psi)=\big(e^{\ve_1}(t+\ve_2),\,e^{-\ve_1}(x+f(t)),\,e^{-\ve_1}(y+\ve_3),\,e^{-3\ve_1}(\psi+g(t)-f_t(t)y)\big).
\end{equation}
In what follows, we set $h(t,y)=g(t)-f_t(t)y$ for convenience and use the substitution $h_y=-f_t$, whenever $h_y$ occurs.

Note that the maximal Lie invariance algebra~$\mathfrak g_0$ of the usual vorticity equation, which is also called the barotropic vorticity equation on the $\ff$-plane and corresponds to the value $\beta=0$,
is much wider than the algebra~$\mathfrak g_1$ and contains~$\mathfrak g_1$ as a proper subalgebra~\cite{andr98Ay,andr88Ay}.
The algebra~$\mathfrak g_0$ is spanned by the vector fields from~$\mathfrak g_1$ jointly with the vector fields
\begin{gather*}
t\p_t-\psi\p_{\psi},\quad
-y\p_x + x\p_y,\quad
-ty\p_x+tx\p_y + \tfrac{1}{2}(x^2+y^2)\p_{\psi},\quad
\tilde h(t)\p_y+\tilde h_t(t)x\p_{\psi},
\end{gather*}
where the parameter $\tilde h$ runs through the set of smooth functions of~$t$.
This means that in addition to the transformations from~$G_1$ the maximal Lie symmetry pseudogroup~$G_0$ of the usual vorticity equation also contains
one more family of scalings, usual rotations in the $(x,y)$-plane, rotations depending on~$t$ with constant angle velocities and generalized Galilean boosts in the $y$-direction.

\begin{remark}\looseness=-1
In order to set up a numerical model, a decision has to be taken about which boundary conditions should be implemented. The numbers of symmetries admitted by a differential equation is almost always reduced for an associated boundary value problem. The most immediate boundary conditions in the atmospheric sciences are periodic ones. However, a periodic domain implies a fixed domain size and therefore breaks the scale invariance of Eq.~\eqref{eq:VorticityEquation}. On the other hand, scale invariance is an equivalence transformation of the class of \emph{all} periodic boundary value problems of the vorticity equation, see also Section~4.1 in~\cite{bihl12By}. A~more serious problem is that the periodicity in $y$-direction is not natural for the beta-plane from the physical point of view. At the same time, using a channel model (rigid walls in the North and in the South of the domain) breaks also the translational invariance in $y$-direction thereby reducing the admitted Lie symmetry group even stronger than in the presence of doubly periodic boundary conditions (though, in contrast to usual hyperdiffusion, it would not be necessary to define an additional boundary condition for the invariant hyperdiffusion as by definition $\psi_x=0$ at the walls of the channel and the diffusion term therefore vanishes there). This is why we will use doubly periodic boundary conditions although $\beta\ne0$ here. Despite this slight inconsistency, doubly periodic boundary conditions are used quite extensively in studying turbulence properties on the beta-plane~\cite{malt91Ay,rhin75Ay,vall06Ay}.
\end{remark}

The above form~\eqref{eq:VorticityEquation} of the vorticity equation is not particularly useful for a numerical evaluation. The reason is, of course, that any numerical model can be run only at a finite resolution, which requires a suitably chosen averaging or filtering of Eq.~\eqref{eq:VorticityEquation}. As from the point of view of invariant parameterization schemes the precise type of averaging is only of secondary importance, we will employ a classical Reynolds averaging to Eq.~\eqref{eq:VorticityEquation} in the paper. This leads to the averaged vorticity equation
\begin{equation}\label{eq:AveragedBetaPlaneEquation}
 \bar \zeta_t+\bar\psi_x\bar\zeta_y-\bar\psi_y\bar\zeta_x+\beta\bar\psi_x = \overline{(\zeta'\psi_y')}_x-\overline{(\zeta'\psi_x')}_y,
\end{equation}
where the right-hand side of this equation denotes the eddy-vorticity flux, which we aim to parameterize subsequently. For the sake of notational simplicity, we will omit bars over the mean quantities from now on.

Slightly more generally, the vorticity equation~\eqref{eq:VorticityEquation} can be augmented with external forcings~$F$ and dissipative terms~$D$ yielding a general expression of the form
\begin{equation}\label{eq:VorticityEquationForcingDamping}
 \zeta_t+\psi_x\zeta_y-\psi_y\zeta_x+\beta\psi_x = F+D.
\end{equation}
A further question we aim to address is whether symmetries might be helpful in deriving invariant expressions for~$F$ and~$D$. As by definition~$F$ denotes \emph{external} forcing terms, it is not immediately clear why symmetries of the vorticity equation should place restrictions on the form of~$F$. However, as we shall show, symmetries are valuable in finding invariant diffusion terms~$D$ that can be used in the course of turbulence modeling. For the sake of simplicity we therefore will use Eq.~\eqref{eq:VorticityEquationForcingDamping} for the case of $F=0$ and $D\ne0$, i.e.\ we assume that no external forcing acts on the system to which a damping is attached. Physically, the presence of~$F$ and~$D$ can be interpreted as symmetry breaking in the vorticity equation~\eqref{eq:VorticityEquation}. Which symmetries are to be broken and which are to be preserved can be controlled upon expressing the term~$D$ via differential invariants derived in Section~\ref{sec:DifferentialInvariantsOfVorticityEquation}. This is a comprehensive problem and not all of the cases arising might be interesting from the physical point of view. We therefore restrict ourselves on the case where~$D$ or the eddy vorticity flux in Eq.~\eqref{eq:AveragedBetaPlaneEquation} can be represented in such a manner that the resulting equation admits all the transformations from the maximal Lie invariance pseudogroup~\eqref{eq:GeneralTransformationVorticityEquation}. This is the approach proposed in~\cite{ober97Ay} and it appears to be suitable for the beta-plane equation.

\section{Algorithm for the construction of differential invariants}\label{sec:AlgorithmOfConstructionOfDiffInvs}

Given a transformation pseudogroup~$G$ in the space of $p$ independent variables $x=(x^1,\dots,x^p)$ and $q$ dependent variables~$u=(u^1,\dots,u^q)$,
the exhaustive description of its differential invariants is reduced to  
either the construction of a functional basis of differential invariants of any fixed order or
finding a complete set of independent operators of invariant differentiation
and a minimal set of differential invariants that generate all differential invariants through invariant differentiation and functional combination~\cite[Section~24]{ovsi82Ay}.
Within the framework of the method of moving frames the solution of this problem is split into two parts \cite{cheh08Ay}.
It is convenient to compute normalized differential invariants and operators of invariant differentiation using the explicit expressions for transformations from~$G$.
The corresponding computation consists of two procedures, \emph{normalization} and \emph{invariantization}.
At the same time, the derivation of syzygies (i.e., relations involving operators of invariant differentiation) between normalized differential invariants
is mostly based on the determining equations of~$G$, and an important tool for this is given by \emph{recurrence formulas}.
In this section we briefly describe related basic notions and results, paying the main attention to the computational realization of algorithms in fixed local coordinates.
See \cite{cheh08Ay,fels98Ay,fels99Ay,olve07Ay,olve08Ay,olve09Cy} for detail and rigorous presentations.

In what follows the index~$j$ runs from 1 to~$p$,
the index~$a$ runs from 1 to~$q$.
We use two kinds of integer tuples for the indexing of objects.
One of these kinds is given by the usual multi-indices of the form $\alpha=(\alpha_1,\dots,\alpha_p)$,
where $\alpha_j\in\mathbb N_0=\mathbb N\cup\{0\}$ and $|\alpha|=\alpha_1+\dots+\alpha_p$.
By $\delta_j$ we denote the $p$-index whose $j$th entry equals 1 and whose other entries are zero.
Thus, both the derivative $\p^{|\alpha|}u^a/(\p x^1)^{\alpha_1}\cdots(\p x^p)^{\alpha_p}$
and the associated variable of the jet space $\mathrm{J}^\infty(x|u)$ are denoted by $u^a_\alpha$,
$\DD^\alpha=\DD_1^{\alpha_1}\cdots\DD_p^{\alpha_p}$, etc.
Here $\DD_j=\p_{x^j}+\sum_{\alpha,a}u^a_{\alpha+\delta_j}\p_{u^a_\alpha}$ is the operator of total differentiation with respect to the variable~$\smash{x^j}$.
The other kind of index tuples is presented by $J=(j_1,\dots,j_\kappa)$,
where $1\leqslant j_k\leqslant p$, $k=1,\dots,\kappa$, $\kappa\in\mathbb N_0$.
Such index tuples are used for the indexing of compositions of operators of invariant differentiation,
which do not commute. Namely, we write $\DD^{\rm i}_J$ for $\DD^{\rm i}_{j_1}\cdots\DD^{\rm i}_{j_\kappa}$.
The symbol~$\dddh$ denotes the horizontal differential,
$\dddh F=\sum_{j=1}^p(\DD_jF)\ddd x^j$ for a differential function~$F=F[u]$,
i.e.\ a function of~$x^j$ and~$u^a_\alpha$.

The normalization procedure for the pseudogroup~$G$ consists of three steps:
\begin{enumerate}
\item
Choose a parameterization (local coordinates) of~$G$ and
find explicit formulas for the prolonged action of~$G$ in terms of the jet variables.
\item
Choose a subset of the transformed jet variables and equate the expressions for them to chosen constants.
\item
Solve the obtained system of normalization equations as a system of algebraic equations with respect to the parameters of the pseudogroup~$G$
including the derivatives of the functional parameters.
\end{enumerate}
The second step is nothing but a choice of an appropriate (coordinate) cross-section of the $G$-orbits.
This should be implemented in a way ensuring that the system from the third step will be well defined and solvable.

The normalization procedure results in the construction of a moving frame $\rho$ for the pseudogroup~$G$,
which is, roughly speaking, an equivariant map from the jet space to~$G$.
Once the moving frame is constructed it can be used to map any object $\chi(x,u_{(n)})$ defined on an open subset of the jet space
(a differential function, a differential operator or a differential form)  to its invariant counterpart, $\iota(\chi(x,u_{(n)}))=\chi(\rho^{(n)}(x,u_{(n)})\cdot (x,u_{(n)}))$.
To carry out this in practice, one should replace all occurrences of the pseudogroup parameters in the transformed version of the object by their expressions obtained with the normalization procedure.

Thus, the invariantization of the coordinate functions~$x^j$ and $u^a_\alpha$ of the jet space yields the so-called \emph{normalized differential invariants} $H^j=\iota(x^j)$ and $I^a_\alpha=\iota(u^a_\alpha)$.
In fact, the invariantized coordinate functions whose transformed counterparts were used to set up the normalization equations are equal to the respective constants chosen in the course of normalization
and hence these objects are called \emph{phantom differential invariants}.
Non-phantom normalized differential invariants are functionally independent and any differential invariant can be represented as a function of normalized differential invariants.
Invariantization of the \emph{operators of total differentiation}, $\DD_j$, gives the operators of invariant differentiation, $\DD^{\rm i}_j$, which upon acting on differential invariants produce other differential invariants.
Note that the domain of the jet space, where invariantized objects are well defined, depends on what cross-section is chosen.

In order to determine the algebra of differential invariants the normalized differential invariants and the operators of invariant differentiation play a key role. It has been proved~\cite{ovsi82Ay} that for any Lie (pseudo)group the algebra of differential invariants can be completely described upon finding a \emph{finite} generating set of differential invariants. As stated above, all the other differential invariants are then a suitable combination of the basis differential invariants or their invariant derivatives. The hardest part in describing the algebra of differential invariants is usually to find a \emph{minimal} generating set of these invariants. Proving the minimality of a given basis usually involves the computation of the \emph{syzygies} among the differential invariants, meaning functional relations among the differentiated differential invariants $\DD^{\rm i}_JI^a_\alpha$, $S(\dots,\DD^{\rm i}_JI^a_\alpha,\dots)=0$.

In general, the normalized differential invariants are derived from invariantization of the derivatives of the dependent variables, whereas the differentiated differential invariants are obtained by acting on normalized differential invariants of lower order with the operators of invariant differentiation. The central point is that the operations of invariant differentiation and invariantization of a differential function in general do not commute. Roughly speaking, the failure of commutation of these two operations is quantified by the so-called recurrence relations
\begin{gather}\label{eq:RecurrenceRelationsCoordinateFunctions}
 \dddh  H^j = \omega^j + \hat\xi^j,\quad \dddh I^a_\alpha = \sum_{j=1}^pI^a_{\alpha+\delta_j}\omega^j+\hat\varphi^{a,\alpha},
\end{gather}
where $\omega^j=\iota(\ddd x^j)$~\cite{cheh08Ay,olve08Ay}.
The forms $\hat\xi^j=\iota(\xi^j)$ and $\hat\varphi^\alpha_a=\iota(\varphi^\alpha_a)$ are the invariantizations of the coefficients of the general prolonged infinitesimal generator
\[
 Q_\infty=\sum_{j=1}^p\xi^j\p_{x^j}+\sum_{\alpha\geqslant0}\sum_{a=1}^q\varphi^{a,\alpha}\p_{u^a_\alpha},\quad \varphi^{a,\alpha}=\DD^\alpha\Bigg(\varphi^a-\sum_{j=1}^p\xi^ju^a_{\delta_j}\Bigg)+\sum_{j=1}^p\xi^ju^a_{\alpha+\delta_j},
\]
of~$G$.
More rigorously, here $\xi^j$ and~$u^a_\alpha$ are interpreted as coordinate functions on the space of prolonged infinitesimal generators of~$G$, i.e., first-order differential forms in the jet space. Hence their invariantizations should also be forms, which are called \emph{invariantized Maurer--Cartan forms}.

The left-hand sides of the relations~\eqref{eq:RecurrenceRelationsCoordinateFunctions} are zero for phantom differential invariants. If the cross-section is chosen in a proper way, the recurrence relations for the phantom invariants can be solved for the independent invariantized Maurer--Cartan forms, which in turn can be plugged into the relations for the non-phantom differential invariants. Collecting coefficients of~$\omega^j$ then yields a closed description of the relation between normalized and differentiated differential invariants, which in turn might enable the determination of a basis of differential invariants. For this latter task, specialized methods from computational algebra can be applied~\cite{olve09Cy}, which is, however, not necessary in the present case due to the relatively simple structure of the maximal Lie invariance pseudogroup~$G_1$ of Eq.~\eqref{eq:VorticityEquation}.

\section{Differential invariants for the beta-plane vorticity equation}\label{sec:DifferentialInvariantsOfVorticityEquation}

In order to derive the moving frame for the maximal Lie invariance pseudogroup~$G_1$ of the barotropic vorticity equation on the beta-plane,
it is necessary to prolong the group actions to the derivatives of $\psi$.
For this aim, we have to derive expressions for the implicit differentiation operators, $\DD_T$, $\DD_X$ and $\DD_Y$.
They can be determined as the dual of the lifted horizontal coframe for~$G_1$, which reads
\begin{align*}
 &\dddh  T=(T_t+\psi_tT_\psi)\ddd t + (T_x+\psi_xT_\psi)\ddd x + (T_y+\psi_yT_\psi)\ddd y = e^{\ve_1}\ddd t \\
 &\dddh  X=(X_t+\psi_tX_\psi)\ddd t + (X_x+\psi_xX_\psi)\ddd x + (X_y+\psi_yX_\psi)\ddd y = e^{-\ve_1}f_t\ddd t+e^{-\ve_1}\ddd x \\
 &\dddh  Y=(Y_t+\psi_tY_\psi)\ddd t + (Y_x+\psi_xY_\psi)\ddd x + (Y_y+\psi_yY_\psi)\ddd y = e^{-\ve_1}\ddd y.
\end{align*}
Therefore, the required implicit differentiation operators are
\begin{equation}\label{eq:ImplicitDifferentiationOperatorsBetaPlaneEquation}
 \DD_T=e^{-\ve_1}(\DD_t-f_t\DD_x),\quad \DD_X=e^{\ve_1}\DD_x,\quad \DD_Y=e^{\ve_1}\DD_y,
\end{equation}
where $\DD_t$, $\DD_x$ and $\DD_y$ denote the usual operators of total differentiation with respect to $t$, $x$ and $y$, respectively,
$\DD_t=\p_t+\sum_{\alpha}\psi_{\alpha+\delta_1}\p_{\psi_\alpha}$,
$\DD_x=\p_x+\sum_{\alpha}\psi_{\alpha+\delta_2}\p_{\psi_\alpha}$ and
$\DD_y=\p_y+\sum_{\alpha}\psi_{\alpha+\delta_3}\p_{\psi_\alpha}$.
Here and in what follows
$\alpha=(\alpha_1,\alpha_2,\alpha_3)$ is a multi-index running through $\mathbb N_0^3$, $|\alpha|=\alpha_1+\alpha_2+\alpha_3$,
$\delta_1=(1,0,0)$, $\delta_2=(0,1,0)$, $\delta_3=(0,0,1)$ and
the variable $\psi_\alpha=\psi_{\alpha_1\alpha_2\alpha_3}$ of the jet space corresponds to the derivative $\p^{|\alpha|}\psi/\p t^{\alpha_1}\p x^{\alpha_2}\p y^{\alpha_3}$.
We also use the notation $f_{(k)}=\ddd^kf/\ddd t^k$ and $h_{(k)}=\p^kh/\p t^k$, $k\in\mathbb N_0$.
The transformed derivatives $\Psi_\alpha=\p^{|\alpha|}\Psi/\p T^{\alpha_1}\p X^{\alpha_2}\p Y^{\alpha_3}$, $|\alpha|>0$, are then
\begin{align*}
\Psi_\alpha&=\DD_T^{\alpha_1}\DD_X^{\alpha_2}\DD_Y^{\alpha_3}\Psi=e^{(\alpha_2+\alpha_3-\alpha_1-3)\varepsilon_1}(\DD_t-f_t\DD_x)^{\alpha_1}\DD_x^{\alpha_2}\DD_y^{\alpha_3}(\psi+h)\\
           &=e^{(\alpha_2+\alpha_3-\alpha_1-3)\varepsilon_1}\left((\DD_t-f_t\DD_x)^{\alpha_1}\psi_{0\alpha_2\alpha_3}
             +\left\{\begin{array}{ll}-f_{(\alpha_1+1)},&\alpha_2=0,\ \alpha_3=1\\ h_{(\alpha_1)},&\alpha_2=\alpha_3=0\end{array}\right\}\right).
\end{align*}

We carry out the normalization procedure in the domain of the jet space which is defined by the condition $\psi_x\ne0$.
We choose the normalization conditions 
\begin{equation}\label{eq:NormalizationBetaPlaneEquation}
T=X=Y=0,\quad \Psi_{k00}=\Psi_{k01}=0,\quad k=0,1,\dots,\quad \Psi_{010}=\varepsilon,
\end{equation}
where $\varepsilon=\sign\psi_x$, which allow us to express all the pseudogroup parameters 
(including the derivatives of functional pseudogroup parameters) in terms of variables of the jet space:
\begin{align}\label{eq:MovingFrameBetaPlaneEquation}
\begin{split}
 &\ve_1=\ln\sqrt{|\psi_x|},\quad \ve_2=-t,\quad \ve_3=-y,\quad f=-x,\\
 &f_{(k+1)}=(\DD_t-\psi_y\DD_x)^k\psi_y, \quad h_{(k)}=-(\DD_t-\psi_y\DD_x)^k\psi, \quad k=0,1,\dots.
\end{split}
\end{align}
In other words, the system~\eqref{eq:MovingFrameBetaPlaneEquation} represents a complete moving frame for the maximal Lie invariance pseudogroup of the vorticity equation.
The series of equalities for $f_{(k+1)}$ an $h_{(k)}$ is proved by induction with respect to~$k$
using the equations \[f_{(k+1)}=(\DD_t-f_t\DD_x)^k\psi_y, \quad h_{(k)}=-(\DD_t-f_t\DD_x)^k\psi.\]

The nontrivial normalized differential invariants are found via invariantizing the derivatives~$\psi_\alpha$ for the values of~$\alpha$
for which $\Psi_\alpha$ are not involved in the construction of the above moving frame, i.e., for
\[
\alpha\in A=\mathbb N_0^3\setminus\{(k,0,0),\,(k,0,1),\,(0,1,0),\,k\in\mathbb N_0\}.
\]
In other words, for each $\alpha\in A$
we should substitute the expressions~\eqref{eq:MovingFrameBetaPlaneEquation} for the pseudogroup parameters into the expressions for~$\Psi_\alpha$.
(The invariantization of the coordinate functions chosen for the normalization conditions~\eqref{eq:NormalizationBetaPlaneEquation} are equal to the corresponding normalization constants and are
the phantom normalized differential invariants for the moving frame~\eqref{eq:MovingFrameBetaPlaneEquation}.)
As a result, we obtain the differential invariants
\begin{gather*}
I_\alpha=\iota(\psi_\alpha)=|\psi_x|^{(\alpha_2+\alpha_3-\alpha_1-3)/2}(\DD_t-\psi_y\DD_x)^{\alpha_1}\psi_{0\alpha_2\alpha_3}, \quad \alpha\in A.
\end{gather*}
The order of $ I_\alpha$ as a differential function of~$\psi$ equals $|\alpha|$.
It is also obvious that any finite number of the invariants~$I_\alpha$ are functionally independent.
This agrees with the general theory of moving frames \cite{cheh08Ay,fels98Ay,olve08Ay}, which also implies a stronger assertion.

\begin{theorem}\label{thm:NormalizedInvsOfBVEGroup}
For each $r\geqslant2$ the functions $I_\alpha=|\psi_x|^{(\alpha_2+\alpha_3-\alpha_1-3)/2}(\DD_t-\psi_y\DD_x)^{\alpha_1}\psi_{0\alpha_2\alpha_3}$,
where $\alpha\in A$ and $|\alpha|\leqslant r$, form a local functional basis of differential invariants of order not greater than~$r$
for the maximal Lie invariance pseudogroup~$G_1$ of the barotropic vorticity equation on the beta-plane.
\end{theorem}

The description of differential invariants of~$G_1$ given in Theorem~\ref{thm:NormalizedInvsOfBVEGroup} is sufficient for applications within the framework of invariant parameterization.
At the same time, it is interesting and useful to have more information on the structure of the algebra of differential invariants of the pseudogroup~$G_1$ including the operators of invariant differentiation.

\begin{theorem}\label{thm:BasisOfDifferentialInvariantsBetaPlaneEquation}
The algebra of differential invariants of the maximal Lie invariance pseudogroup
of the barotropic vorticity equation on the beta-plane~\eqref{eq:VorticityEquation}
is generated, in the domain~$\Omega_1$  of the jet space where $\DD_x^{\,2}(\sqrt{|\psi_x|}\,)\ne0$,
by the single differential invariant $I_{020}=\psi_{xx}/\sqrt{|\psi_x|}$
along with the three operators of invariant differentiation
\[
  \DD^{\rm i}_t=\frac{1}{\sqrt{|\psi_x|}}(\DD_t-\psi_y\DD_x),\quad \DD^{\rm i}_x=\sqrt{|\psi_x|}\DD_x,\quad \DD^{\rm i}_y=\sqrt{|\psi_x|}\DD_y.
\]
\end{theorem}

All other differential invariants are functions of $I_{020}$ and invariant derivatives thereof. The proof of this theorem is presented in detail in Appendix~\ref{sec:AppendixStructureAlgebraDifferentialInvariants}.

\section{Invariantization of parameterization schemes}\label{sec:InvariantizationOfBetaPlaneBVE}

The Replacement Theorem states that any differential invariant $I(x,u_{(n)})$ of order~$n$ can be expressed in terms of the normalized differential invariants
via replacing any argument of~$I(x,u_{(n)})$ by its respective invariantization, see~\cite{fels99Ay}.
In particular, any system of differential equations can be represented using the normalized differential invariants of its associated maximal Lie invariance group.
The invariantization of the vorticity equation~\eqref{eq:VorticityEquation} in view of the moving frame~\eqref{eq:MovingFrameBetaPlaneEquation} reads
$
(I_{120}+I_{102}) + (I_{021} + I_{003}) + \beta = 0,
$
or, explicitly
\begin{equation}\label{eq:InvariantRepresentationBetaPlaneEquation}
\frac{\zeta_t-\psi_y\zeta_x}{\psi_x} + \zeta_y + \beta = 0.
\end{equation}
This is the fully invariant representation of the barotropic vorticity equation on the beta-plane.

Differential invariants computed in the previous section can be assembled together to invariant parameterizations of the eddy-vorticity flux in the averaged vorticity equation~\eqref{eq:AveragedBetaPlaneEquation}. Alternatively, we can invariantize any existing parameterization scheme under the moving frame action~\eqref{eq:MovingFrameBetaPlaneEquation}. The following two examples implement this idea.

\begin{example}
A classical albeit simple parameterization for the eddy-vorticity flux is
\[
{\rm evf}:=\overline{(\zeta'\psi_y')}_x-\overline{(\zeta'\psi_x')}_y=\DD_x(K\zeta_x)+\DD_y(K\zeta_y),
\]
where $K=K(x,y)$ might be considered as a spatially dependent function. The most straightforward way to cast this parameterization into the related invariant one is by applying the moving frame~\eqref{eq:MovingFrameBetaPlaneEquation} to the terms on the right-hand side. This yields
\[
\begin{split}
{\rm evf}^{\rm i}&=\DD^{\rm i}_x(K(I_{030}+I_{012}))+\DD^{\rm i}_y(K(I_{021}+I_{003}))=K(I_{040}+2I_{022}+I_{004})\\
                 &=K\sqrt{|\psi_x|}(\zeta_{xx}+\zeta_{yy}),
\end{split}
\]
where ${\rm evf}^{\rm i}=\iota({\rm evf})$ and $K=\const$ now as $\iota(x)=\iota(y)=0$. The invariant representation of the closed barotropic vorticity equation then reads
 \[
  \frac{\zeta_t-\psi_y\zeta_x}{\psi_x} + \zeta_y + \beta = K\sqrt{|\psi_x|}(\zeta_{xx}+\zeta_{yy}).
 \]
\end{example}

\begin{example}
The anticipated (potential) vorticity method was originally proposed by Sadourny and Basdevant~\cite{sado85Ay}. The idea of this method is to approximate the diffusion effect in the vorticity equation by a weighted upwind estimate of the vorticity itself, i.e.\ by employing
\[
 \zeta^{\rm a}_t + J(\psi,\zeta^{\rm a}) = \nu J(\psi,\Delta^nJ(\psi,\zeta^{\rm a})),
\]
where $\nu$ is a constant, $n\in\mathbb N_0$ and $\zeta^{\rm a}$ is the absolute vorticity. Here and in what follows $\Delta=\nn^2$ is the two-dimensional Laplacian. The purpose of adding the specific forcing term on the right-hand side of the vorticity equation is to suppress the high frequency noise in the vorticity field and at the same time to ensure that energy is conserved during the integration while enstrophy is dissipated. The latter properties can be easily verified upon multiplying Eq.~\eqref{eq:VorticityEquation} with the stream function~$\psi$ and any function of the absolute vorticity~$\zeta^{\rm a}$, respectively, and integrating over the domain~$\Omega$, see also~\cite{vall88Ay}.

There is a problem with this parameterization scheme in that it is not Galilean invariant. Galilean invariance (as well as the proper scale invariance), however, can be easily included by the method of invariantization. For the sake of demonstration, we consider the case of $n=0$ here, which is the original version of the anticipated vorticity closure. Upon using the moving frame~\eqref{eq:MovingFrameBetaPlaneEquation}, we obtain
\[
\iota(J(\psi,J(\psi,\zeta^{\rm a})))=\frac{1}{\sqrt{|\psi_x|}}J(\psi_y,\zeta^{\rm a})+\sqrt{|\psi_x|}\zeta^{\rm a}_{yy}.
\]
Attaching this to the invariant representation of the vorticity equation~\eqref{eq:InvariantRepresentationBetaPlaneEquation}, the vorticity equation with fully invariant closure reads ($\varepsilon=\sign\psi_x$)
\begin{equation}\label{eq:VorticityEquationWithAnticipatedMethod}
\zeta^{\rm a}_t+J(\psi,\zeta^{\rm a})=\nu\sqrt{|\psi_x|}(\varepsilon J(\psi_y,\zeta^{\rm a})+\psi_x\zeta^{\rm a}_{yy}).
\end{equation}
It is obvious that this parameterization is quite different from that proposed in~\cite{sado85Ay}. It cannot be brought in the form of nested Jacobian operators and it does not conserve energy any more (for the derivation of conservative invariant closure schemes, see Section~\ref{SectionOnConservativeInvariantParameterization}). On the other hand, the inherent invariance of the closed vorticity equation~\eqref{eq:VorticityEquationWithAnticipatedMethod} with respect to Galilean and scale symmetries is an appealing property for itself and might be relevant e.g.\ when vorticity dynamics is studied in a moving coordinate frame.

Quite recently, an approximate scale invariant formulation of the anticipated potential vorticity method was proposed in~\cite{chen11Ay} using scale analysis techniques and physical reasoning. The motivation for this study is that modern weather and climate models might be required to operate on grids with variable resolution. Unfortunately, varying resolution in an atmospheric numerical model is not a simple task as most of the parameterization schemes employed are definitely not scale invariant, but rather tuned to yield best results on a fixed grid. Painful efforts might be necessary in order to adjust parameterization schemes of a numerical model to various spatial-temporal resolutions. Having a general method for deriving scale-insensitive closure schemes at hand is therefore of potential practical interest in numerical geophysical fluid dynamics. Albeit simple, the method of invariantization of existing parameterization schemes may give appropriate closure schemes that are both physically meaningful and respect essential symmetries of a specific process to be represented numerically.
\end{example}

These are only two examples for fully invariant closure schemes.
See one more example in the next section.
In principle, each term of the form $S(I^1,\dots,I^N)$,
where $S$ is a smooth function of its arguments
and $I^1$, \dots, $I^N$ are differential invariants of~$G_1$,
satisfies the same requirement when added to the right hand side of Eq.~\eqref{eq:InvariantRepresentationBetaPlaneEquation}.
In other words, the general form of closure ansatzes for Eq.~\eqref{eq:InvariantRepresentationBetaPlaneEquation}, which are invariant with respect to the entire group~$G_1$, is
\[
\zeta_t+\psi_x\zeta_y-\psi_y\zeta_x+\beta\psi_x=\psi_xS(I^1,\dots,I^N).
\]

\section{Application of invariant parameterizations to \texorpdfstring{\\}{ } turbulence modeling}\label{sec:ApplicationOfInvariantParameterization}

In this section, we give an application in which we aim to demonstrate in practice the ideas outlined above and in~\cite{popo10Cy}. This example deals with turbulence properties of the two-dimensional incompressible Euler equations. Strictly speaking, turbulence is a three-dimensional problem as a two-dimensional turbulent flow is not stable with respect to fully three-dimensional perturbations to that flow~\cite{salm98Ay}. Nevertheless, there are countless studies concerning the turbulent properties of two-dimensional flow simply because it is a relevant problem in large-scale geophysical fluid dynamics, which behaves as approximately two-dimensional.

In short, the first theoretical results concerning two-dimensional turbulence were derived in~\cite{batc69Ay,krai67Ay}, following the pioneering work on three-dimensional turbulence done by Kolmogorov~\cite{kolm41Ay}. Extensive numerical studies have been carried out since then attempting to verify distinct aspects of the theory proposed \cite{benz86Ay,benz88Ay,brac00Ay,fox10Ay,lill71Ay}. The two-dimensional case is especially peculiar, as it admits infinitely many conservation laws including the conservation of energy. The energy in the barotropic vorticity is purely kinetic and can be represented in different ways using doubly periodic boundary conditions as
\begin{equation}\label{eq:EnergyVorticityEquation}
 \mathcal E = \frac12\int_\Omega\vv^2\ddd A = \frac12\int_\Omega(\nn\psi)^2\ddd A = -\frac12\int_\Omega\psi\zeta\ddd A,
\end{equation}
where $\Omega = [0, L_x[{}\times[0,L_y[$ and $\ddd A = \ddd x\,\ddd y$. The special form of Eq.~\eqref{eq:VorticityEquation} leads to the following class of conservation laws
\[
 \mathcal C_g = \int_\Omega g(\zeta^{\rm a})\ddd A,
\]
for any smooth function~$g$ of the absolute vorticity $\zeta^{\rm a}=\zeta+\ff_0+\beta y$.
The most relevant realization of the above conservation laws in the present context is the enstrophy, given for the particular value~$g=(\zeta^{\rm a})^2/2$.

First of all, consider the case of no differential rotation ($\beta=0$), i.e.\ the Coriolis parameter~$\ff$ is approximated by the constant $\ff_0$, which is referred to as the $\ff$-plane approximation. It is the simultaneous conservation of energy and enstrophy in this case that leads to the remarkable behavior of two-dimensional turbulence~\cite{salm98Ay,vall06Ay}. Starting with a random initial velocity (or stream function field), energy is transported to the large scale, while enstrophy is transported to the smaller scales. This cascade is associated with an organization of the vortices, with vortices of the same sign merging into bigger ones (though the precise mechanisms of the cascade including the role of the vortices are not yet fully understood). In order to initiate these fluxes of energy to the larger scale and enstrophy to the smaller scale and thus the process of organization, it is necessary to place a sink of enstrophy at the very small scales. This sink acts as a remover of enstrophy while ideally conserving energy, as the latter is transported away from the small scales on which the dissipation acts (which in practice is hard to realize in a numerical simulation using a finite number of grid points). It is believed that the form of the energy spectrum in a range above which dissipation is acting (inertial range) can be derived using scaling theory in a similar manner as it was shown by Kolmogorov for the three-dimensional case~\cite{salm98Ay,vall06Ay}.

The energy and enstrophy spectra $E(k)$ and $C(k)$ are defined by
\begin{align*}
 &\bar{\mathcal E}=\frac{1}{2L_xL_y}\int_\Omega\vv^2\ddd A=\frac{1}{2L_xL_y}\int_\Omega(\nn\psi)^2\ddd A=\int E(k)\ddd k,\\[1ex]
 &\bar{\mathcal C}=\frac{1}{2L_xL_y}\int_\Omega\zeta^2\ddd A=\frac{1}{2L_xL_y}\int_\Omega(\Delta\psi)^2\ddd A=\int C(k)\ddd k,
\end{align*}
where $\bar{\mathcal E}$ and $\bar{\mathcal C}$ are the average energy and average enstrophy, $k=\sqrt{(k^x)^2+(k^y)^2}$ is the scalar wave number, $k^x$ and $k^y$ are the wave numbers in $x$- and $y$-direction, respectively. The possibility of using a single wave number is due to the assumption of isotropy that is generally made in turbulence theory and which is reasonable in the case of vanishing differential rotation~\cite{vall06Ay}. According to the theory, the form of the energy spectrum in the inertial range should follow
\[
E(k)\propto k^{-3}.
\]
This is referred to as the \emph{enstrophy cascade} in two-dimensional turbulence. Analogously, the \textit{enstrophy spectrum} in the inertial range should follow
\[
 C_{\rm ens}(k)\propto k^{-1}=k^2E(k).
\]

The impact of the beta-term in the vorticity equation on the turbulent cascades was first studied in~\cite{rhin75Ay}. In this seminal paper, it was remarked that the Rossby wave solutions admitted by the beta-plane equation can act as a source of anisotropization of turbulence at the larger scale. Qualitatively, at some stage the size of the vortices is big enough that they are exposed to the effect of differential rotation, which essentially hinders the tendency of vortex growth due to the inverse energy cascade. Rather, the vortices evolve into Rossby wave and eventually to the formation of zonal jets as observed e.g.\ on giant planets. Depending on the precise setting used (e.g.\ strength of the differential rotation, additional energy injection to the system), the results of turbulence simulations can vary, but often energy spectra steeper than those predicted theoretically can be found~\cite{huan01Ay,malt91Ay,rhin75Ay}.

In practice, the sink of enstrophy at the small scales is usually implemented by adding a hyperviscosity of the form
\begin{equation}\label{eq:non-invariantHyperdiffusion}
   D = (-1)^{n-1}\nu\Delta^n\zeta
\end{equation}
for $n \in \mathbb{N}^+$ to the right-hand side of Eq.~\eqref{eq:VorticityEquation}, cf.\ Eq.~\eqref{eq:VorticityEquationForcingDamping}. However, it can easily be checked that this form of hyperviscosity is not invariant under the Lie symmetry pseudogroups of the beta-plane and $\ff$-plane equations. More specifically, it violates the scale invariance of Eq.~\eqref{eq:VorticityEquation}. \emph{From the theoretical point of view, this violation appears to be especially odd, as it is precisely the scale invariance of the Euler equations that is used to derive the form of the energy spectrum in the inertial range.}

Theorem~\ref{thm:NormalizedInvsOfBVEGroup} directly implies that the invariantization
$\iota(D) = (-1)^{n-1}\nu\sqrt{|\psi_x|^{2n-1}}\Delta^n\zeta$
is a differential invariant of the maximal Lie invariance pseudogroup of the vorticity equation.
In view of the results of Section~\ref{sec:InvariantizationOfBetaPlaneBVE},
we conclude that the form of the diffusion term obtained in the course of the invariantization is
\[
\tilde D=|\psi_x|\iota(D)=(-1)^{n-1}\nu\sqrt{|\psi_x|^{2n+1}}\Delta^n\zeta.
\]

The completely invariant formulation of the vorticity equation on the beta-plane with hyperdiffusion therefore reads
\begin{equation}\label{eq:BetaPlaneEquationInvariantHyperdiffusion}
 \zeta_t + \psi_x\zeta_y-\psi_y\zeta_x+\beta\psi_x = (-1)^{n-1}\nu\sqrt{|\psi_x|^{2n+1}}\Delta^n\zeta.
\end{equation}
Note, however, that the price for introducing an invariant enstrophy sink is the \emph{nonlinearity} of the (hyper)diffusion term. More generally, the situation is alike to the problem of finding a relation between the Reynolds stresses and the mean strain rate in the Reynolds averaged Navier--Stokes equations or in large--eddy simulations thereof. It was pointed out that establishing a relationship between the \emph{nonlinear} Reynolds stresses and the \emph{linear} strain rate (i.e.\ invoking the Boussinesq hypothesis) may lead to unrealistic results for certain turbulent flows such as in rotating or stratified fluids or those exposed to abrupt changes of the mean strain rate, see the discussions in~\cite{pope00Ay,wilc93Ay}. It is therefore worthwhile pointing out that the requirement of preserving the entire maximal Lie invariance pseudogroup of the barotropic vorticity equation on the beta-plane automatically yields nonlinear hyperdiffusion terms. For $n=1$, the right-hand side of Eq.~\eqref{eq:BetaPlaneEquationInvariantHyperdiffusion} can be considered as a generalized down-gradient parameterization for the eddy-vorticity flux, which is also a nonlinear quantity. That is, requiring a (hyper)diffusion scheme to be scale invariant, it is indispensable to use \emph{nonlinear} (hyper)diffusion.

It is important to note that the anisotropic coefficient $\sqrt{|\psi_x|^{2n+1}}$ arises due to the special form of normalization conditions~\eqref{eq:NormalizationBetaPlaneEquation} we have chosen in Section~\ref{sec:DifferentialInvariantsOfVorticityEquation} for the construction of the moving frame. This form is by no means unique but rather a consequence of the moving frame we have invoked. The situation is comparable to the discretization of differential equations, which can also be done in multiple ways. Some schemes have better properties than others and ultimately it is necessary to both analyze and test the various schemes for different sets of problems. Having more than one possibility to construct invariant subgrid-scale schemes out of a given non-invariant scheme should therefore be considered as an advantage rather than as a drawback of the proposed method. The knowledge of the complete set of differential invariants, which is obtained as a byproduct when determining the invariantization map for a given group action, allows one to derive series of invariant closure schemes starting from that obtained as a direct result of the invariantization of the given initial scheme. This is facilitated by recombining a given invariant scheme using the differential invariants, as any functional combination of differential invariants is again a differential \mbox{invariant}.
\looseness=1

A number of alternative (isotropic) forms of a completely invariant nonlinear hyperviscosity term for the vorticity equation on the beta-plane can therefore be suggested, e.g.
\[
\tilde D=(-1)^{n-1}\nu\zeta^{2n+1}\Delta^n\zeta,\quad
\tilde D=(-1)^{n-1}\nu\nn(\zeta^{2n+1}\nn\Delta^{n-1}\zeta),\quad\mbox{etc.},
\]
which are derived upon recombining the differential invariants derived in Theorem~\ref{thm:NormalizedInvsOfBVEGroup}.
Due to the wide possibility for varying ansatzes for invariant parameterizations we can control different desirable conditions which proper invariant closure schemes should additionally satisfy,
cf.\ Section~\ref{SectionOnConservativeInvariantParameterization}.

Subsequently we will exclusively work with Eq.~\eqref{eq:BetaPlaneEquationInvariantHyperdiffusion}. Our motivation for choosing the anisotropic hyperdiffusion~\eqref{eq:BetaPlaneEquationInvariantHyperdiffusion} rather than any of the above isotropic ones stems from recent experiments on turbulence which suggest that contrary to the Kolmogorov hypothesis the small scales might indeed feel the effects from the large scale being anisotropic, i.e.\ that anisotropy can propagate through to the very small scales, see e.g.~\cite{shen00Ay}. However, future tests will be conducted so as to compare the different forms of invariant hyperdiffusion.

We give some numerical experiments using Eq.~\eqref{eq:BetaPlaneEquationInvariantHyperdiffusion} and compare it with the respective non-invariant model that employs classical hyperdiffusion~\eqref{eq:non-invariantHyperdiffusion}. Both models are integrated using a finite difference scheme and biharmonic dissipation is used in all the experiments, i.e.\ $n=2$. The nonlinear terms on the left-hand side are discretized using the Arakawa Jacobian operator~\cite{arak66Ay}, which guarantees energy and enstrophy conservation of the spatial discretization in the case of vanishing dissipation, $\nu=0$. A leapfrog scheme is used for the time stepping in conjunction with a Robert--Asselin--Williams filter~\cite{will09Ay}, in order to suppress the computational mode. The size of the domain is $L_x=L_y=2\pi$, with a default of $N=1024$ grid points in each direction, $\beta=1$. The initial condition is a Gaussian random stream function field, with the initial energy spectrum given by the function $E(k)\propto k^{3}\exp(-3k^2/k_p^2)$, where $k_p=64$. No normalization of the initial energy was used. The value of $\nu$ was chosen to be $\nu_{\rm inv}=1\cdot10^{-10}$ in the invariant case and $\nu_{\rm ninv}=2\cdot 10^{-9}$ for the non-invariant simulations. Note that the value of $\nu_{\rm ninv}$ has been selected to lie in between the values given in~\cite{brac00Ay} for the two integrations using $512^2$ and $4096^2$ grid points. The value of $\nu_{\rm inv}$ has been chosen so that $\nu_{\rm inv}\approx {\rm max}(\nu_{\rm ninv}\sqrt{|\psi_x|^{5}})$ initially for the sake of comparison.

Both models have been integrated for approximately 10 000 time steps using $\Delta t=1\cdot 10^{-3}$. Hence, all the results presented below were evaluated at approximately $t=10$, which should be long enough so that inertial ranges can form in the energy and enstrophy spectra. Below, we shall like to present the enstrophy spectra for fully developed freely decaying turbulence using both the invariant and the non-invariant hyperdiffusion terms. As was said above, according to the Batchelor--Kraichnan theory the enstrophy spectrum should be of the form $k^{-1}$ in the inertial range. However, finding experimental evidence for a spectrum of this form proved rather hard and most numerical simulations carried out so far yield steeper \mbox{spectra}.
\looseness=1

\begin{figure}[!t]
	\centering
	\includegraphics{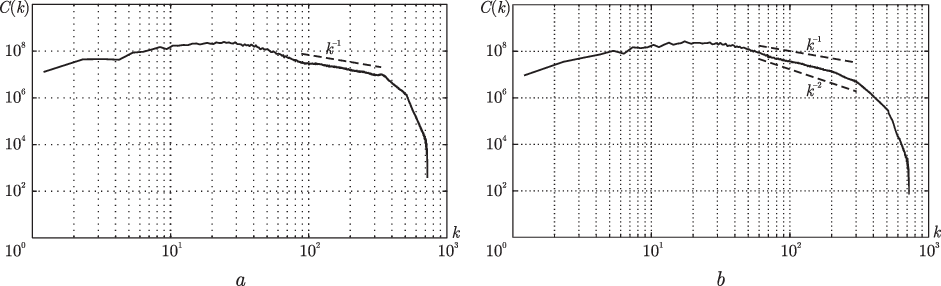}
	\caption{Enstrophy spectrum at approximately $t=10$ using ({\it a}) invariant hyperdiffusion and ({\it b}) non-invariant hyperdiffusion.}
	\label{fig:EnstrophySpectrum}
\end{figure}

\begin{figure}[!t]
	\centering
	\includegraphics{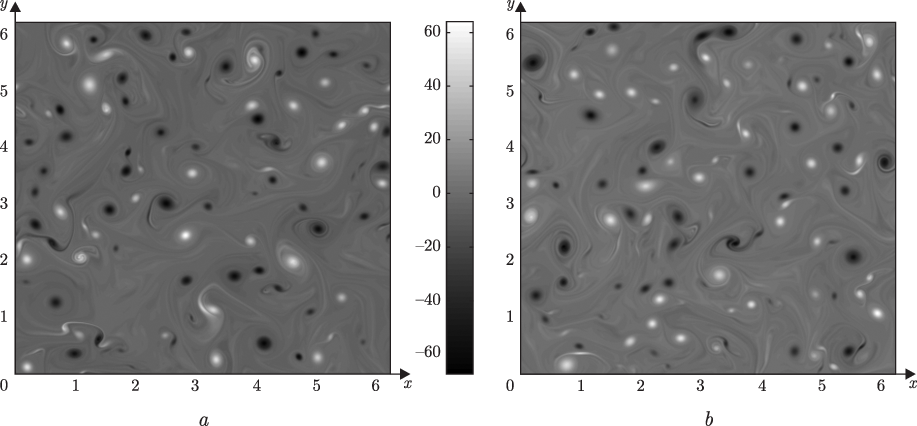}
	\caption{Vorticity field at approximately $t=10$ using ({\it a}) invariant hyperdiffusion and ({\it b}) non-invariant hyperdiffusion.}
	\label{fig:VorticityField}
\end{figure}

In Fig.~\ref{fig:EnstrophySpectrum}{\it a} we show the enstrophy spectrum found from the simulation using invariant hyperdiffusion. In the region between approximately $k=100$ up to $k=300$ the spectrum follows $k^{-1}$ almost perfectly. That is, the invariant hyperdiffusion of the form used in~\eqref{eq:BetaPlaneEquationInvariantHyperdiffusion} leads to an experimental verification of the Batchelor--Kraichnan theory.

In Fig.~\ref{fig:EnstrophySpectrum}{\it b} we show the corresponding enstrophy spectrum obtained using conventional (non-invariant) hyperdiffusion. As in the majority of turbulence simulations, also we obtain a spectrum in the inertial range that is \textit{steeper} than $k^{-1}$, lying between $k^{-1}$ and $k^{-2}$, in this case. Moreover, it is instructive to note that the lower parts of the spectra (up to the respective inertial ranges) are rather similar for both schemes, while differences occur within the inertial and in the diffusion ranges. This observation underpins that the proposed nonlinear invariant hyperdiffusion is physically acting as a viscosity term in Eq.~\eqref{eq:BetaPlaneEquationInvariantHyperdiffusion}.

Fig.~\ref{fig:VorticityField} shows the associated vorticity fields obtained using the invariant and non-invariant hyperdiffusion schemes at the end of the integration. Note that the value of $\beta$ chosen is rather small (and much smaller as compared to the value of $\beta=3$ used in~\cite{malt91Ay}) so the effects of differential rotation on the vorticity fields are rather minimal. Both fields look qualitatively similar verifying that invariant hyperdiffusion is capable of producing a physically meaningful vorticity field.

\begin{remark}
 Decaying turbulence simulations are an important class of tests for numerical integration schemes. On the other hand, from the point of view of both the theory and application, it is generally more instructive when Eq.~\eqref{eq:VorticityEquation} is augmented with some forcing which supplies energy to the system and thereby prevents turbulence from dying out. As it is then usually necessary to damp out the energy which is otherwise piling up at small wave numbers (large scales) due to the inverse energy cascade, an additional drag term is introduced in Eq.~\eqref{eq:VorticityEquation}. This drag term can be either physical (e.g.\ linear Ekman drag due to bottom friction) or, similar as hyperviscosity, scale selective. In the latter case, one uses a \emph{hypoviscosity}~\cite{dani01Ay}, which is given by adding a term proportional to $\Delta^{-n}\zeta$, which acts scale selective by emphasizing the large scale and thus is effectively energy removing. Again, one could raise the question whether such a hypofriction should possess some invariance properties, but this is beyond the scope of the present paper and should be considered in a forthcoming study.
\end{remark}

\section{Conservative invariant parameterizations}\label{SectionOnConservativeInvariantParameterization}

A parameterization is called \emph{conservative} if the corresponding closed system of differential equations possesses nonzero conservation laws.
Special attention should be paid to parameterizations possessing conservation laws that have a clear physical interpretation (such as the conservation of energy, mass and momentum)
and that originate from the conservation laws of the initial system of equations.
If a parameterization is both conservative and invariant with respect to a pseudogroup of transformations,
it is called a \emph{conservative invariant parameterization}.

The general method for singling out conservative parameterizations among invariant closure ansatzes is based on the usage of the Euler operators,
i.e.\ variational derivatives with respect to the dependent variables~\cite{olve86Ay}.
Suppose that $\tilde{\mathcal L}_\theta$: $\tilde L^l(x,\bar u_{(n)},\theta)=0$, $l=1,\dots,m$, $\theta=\theta(I^1,\dots,I^N)$
represent a family of local parameterizations for a system~$\mathcal L$: $L^l(x,u_{(n)})=0$, $l=1,\dots,m$,
which are invariant with respect to a pseudogroup~$G$.
Here $\tilde L^l$ are fixed smooth functions of their arguments.
The tuple~$\theta$ of arbitrary elements consists of smooth functions of certain differential invariants $I^1$, \dots, $I^N$ of~$G$.
It runs through a set of such tuples constrained by a system of differential equations, where $I^1$, \dots, $I^N$ play the role of independent variables.
We require the tuples $(\lambda^{m1},\dots,\lambda^{ml})$, $m=1,\dots,M$, of differential functions of~$u$
to be characteristics of $M$ linearly independent local conservation laws of the system~$\tilde{\mathcal L}_\theta$ for some values of~$\theta$,
i.e.\ for each~$m$ the combination $\lambda^{m1}\tilde L^1+\dots+\lambda^{ml}\tilde L^l$ is a total divergence.
The theorem on characterization of total divergences~\cite[Theorem~4.7]{olve86Ay} then implies the equations
\mbox{$
\mathop{{\sf E}^a}\nolimits(\lambda^{m1}\tilde L^1+\dots+\lambda^{ml}\tilde L^l)=0
$}
for each~$m=1,\dots,M$ and $a=1,\dots,q$, 
where $\mathop{{\sf E}^a}\nolimits$ is the Euler operator associated with the dependent variable~$u^a$, 
${\sf E}^a f=\sum_{\alpha}(-\DD)^\alpha f_{u^a_\alpha}$.
Splitting these equations with respect to derivatives of~$u$ wherever this is possible,
one constructs the system of determining equations with respect to~$\theta$, which should be solved
in order to derive the corresponding conservative invariant parameterizations.

As the direct computation is too cumbersome, we use some heuristic arguments and
look for a diffusion ansatz for the barotropic vorticity equation on the beta-plane that satisfies the following relevant and valuable conditions:
\begin{itemize}\itemsep=0ex
\item
It is invariant with respect to the entire maximal Lie invariance pseudogroup~$G_1$ of Eq.~\eqref{eq:VorticityEquation}.
\item
The subgrid-scale term or, more generally, the sink term to be represented is a differential function of the vorticity~$\zeta$ 
(namely, a polynomial depending only on derivatives of~$\zeta$ with respect to the space variables~$x$ and~$y$).
\item
This expression is as similar as possible to the hyperviscosity term~\eqref{eq:non-invariantHyperdiffusion}.
\item
And, last but not least, the parameterization is conservative. More precisely, it possesses all the conservation laws of Eq.~\eqref{eq:VorticityEquation} with zero-order characteristics.
\end{itemize}
The second point guarantees the invariance of the corresponding diffusion ansatz under all transformations from~$G_1$ that do not involve scalings. 
In order to provide the scale invariance, we should just balance the scaling weights of derivatives of~$\zeta$ in the diffusion term. 
Moreover, these derivatives should be composed in such a way that allows integrating by parts 
in order to represent the diffusion term multiplied by an arbitrary zero-order conservation-law characteristic of Eq.~\eqref{eq:VorticityEquation}
in conserved form.
An example of such a parameterization is given by
\begin{equation}\label{eq:RemarkableParameterization}
\zeta_t+\psi_x\zeta_y-\psi_y\zeta_x+\beta\psi_x=D, \quad D=\nu\Delta\frac{\Delta\zeta^7}\zeta=7\nu\Delta(\zeta^5\Delta\zeta+6\zeta^4(\nn\zeta)^2).
\end{equation}

All the properties listed above can be checked for the sink term~\eqref{eq:RemarkableParameterization}.
Thus, the expression for~$D$ from~\eqref{eq:RemarkableParameterization} involves only the vorticity and its derivatives and is quite similar to~\eqref{eq:non-invariantHyperdiffusion}.
Moreover, the diffusion~$D$ is a globally defined differential function which is a polynomial of its arguments.
The invariance of Eq.~\eqref{eq:RemarkableParameterization} with respect to~$G_1$ can be simply checked using the infinitesimal invariance criterion.
A more sophisticated way to check this invariance is to rewrite Eq.~\eqref{eq:RemarkableParameterization} in terms of normalized invariants of the pseudogroup~$G_1$,
which will not be done explicitly here.
As an unexpected but valuable bonus we have that the maximal Lie symmetry pseudogroup of Eq.~\eqref{eq:RemarkableParameterization} with the same term~$D$ in the case of the $\ff$-plane ($\beta=0$) is even wider than~$G_1$.
It also includes the usual rotations of the variables~$(x,y)$ and the generalized Galilean boosts in $y$-direction, which belong to
the Lie symmetry pseudogroup~$G_0$ of the barotropic vorticity equation on the $\ff$-plane.
This in particular means that the parameterization~\eqref{eq:RemarkableParameterization} is isotropic.

The space of zero-order characteristics of Eq.~\eqref{eq:VorticityEquation} is generated by the characteristics
$\lambda=f(t)$, $\lambda=g(t)y$ and $\lambda=\psi$, where~$f$ and~$g$ run through the set of smooth functions of~$t$.
The physically most important of these characteristics are $\lambda=1$, $\lambda=y$ and $\lambda=\psi$, which are associated with the conservation of circulation, $x$-momentum and energy.
Any zero-order characteristic of Eq.~\eqref{eq:VorticityEquation} is a characteristic of Eq.~\eqref{eq:RemarkableParameterization}.
Indeed, denoting \[L:=\zeta_t+\psi_x\zeta_y-\psi_y\zeta_x+\beta\psi_x-D\] we derive that
\begin{align*}
fL={}&\DD_x\left(f\psi_{xt}+f\psi\zeta_y+f\beta\psi-\nu f\DD_x\frac{\Delta\zeta^7}\zeta\right)
+\DD_y\left(f\psi_{yt}-f\psi\zeta_x-\nu f\DD_y\frac{\Delta\zeta^7}\zeta\right),
\\
gyL={}&\DD_x\left(gy\psi_{xt}+gy\psi\zeta_y-\frac g2(\psi_y)^2+gy\beta\psi-\nu gy\DD_x\frac{\Delta\zeta^7}\zeta\right)\\
&+\DD_y\left(gy\psi_{yt}-g\psi_y-gy\psi\zeta_x+g\psi\psi_{xy}-\nu gy\DD_y\frac{\Delta\zeta^7}\zeta+\nu g\frac{\Delta\zeta^7}\zeta\right),
\\
\psi L={}&
 \DD_t\left(-\frac12(\nn\psi)^2\right)
+\DD_x\left(\psi\psi_{xt}+\frac12\psi^2\zeta_y+\frac\beta2\psi^2-\nu\psi\DD_x\frac{\Delta\zeta^7}\zeta+\nu\psi_x\frac{\Delta\zeta^7}\zeta-\nu\DD_x\zeta^7\right)\\
&+\DD_y\left(\psi\psi_{yt}-\frac12\psi^2\zeta_x-\nu\psi\DD_y\frac{\Delta\zeta^7}\zeta+\nu\psi_y\frac{\Delta\zeta^7}\zeta-\nu\DD_y\zeta^7\right).
\end{align*}
If we grant that the vorticity equation coupled with some diffusive term possesses a smaller number of conservation laws (e.g.\ owing to the special physical properties of this diffusion),
we can use a simpler form for the expression~$D$.
For example, the differential function $D=\nu\Delta\zeta^4$ leads to a parameterization which is invariant with respect to the entire pseudogroup~$G_1$ and
possesses conservation laws with the characteristics $\lambda=f(t)$, $\lambda=g(t)y$ for arbitrary values of the smooth parameter-functions~$f$ and~$g$.

The parameterization~\eqref{eq:RemarkableParameterization} demonstrates the feasibility of combining invariant and conservative properties of closure schemes. This possibility is important for two obvious reasons. Firstly, conservation laws incorporate relevant physical information that is worth being preserved by a parameterization scheme. Secondly, from the point of view of constructing parameterization schemes, the requirement of preserving both symmetries and conservation laws leads to a more specific class of schemes than considering either only symmetries or only conservation laws. The additional narrowing of the class of admitted schemes using geometric constraints can then help to reduce the number of schemes that must be tested numerically so as to find the optimal parameterization for a given process.

\section{Conclusion and discussion}\label{sec:Conclusion}

The differential invariants of the Lie symmetry pseudogroup~$G_1$ of the barotropic vorticity equation on the beta-plane are computed using the technique of moving frames for Lie pseudogroups. A basis of these differential invariants along with the associated operators of invariant differentiation is established. Together, they serve to completely describe the algebra of differential invariants of~$G_1$. Although differential invariants have many applications (such as the integration of ordinary differential equations~\cite{olve86Ay}, computation of so-called differentially invariant solutions~\cite{golo04Ay,ovsi82Ay} and the construction of invariant numerical discretization schemes~\cite{doro11Ay}), in the paper we focus on their usage in the construction of invariant closure schemes or, perhaps more generally, invariant diffusion terms for the averaged vorticity equation. This is one of the two general methods proposed in~\cite{popo10Cy} to derive parameterization schemes with symmetry properties. As an alternative to the direct usage of elementary differential invariants that can be build together to yield invariant closure schemes, we propose the method of invariantization of existing parameterization schemes. This method is along the line of invariantization of existing discretization schemes as introduced in~\cite{kim06Ay,kim08Ay}. Although this method is straightforward to apply, a potential complication is that the result depends on the particular choice of the moving frame and therefore does not lead to a unique invariant counterpart of an existing non-invariant scheme. As a consequence, it might be necessary to modify invariantized closure schemes and to test different invariantizations in order to devise physically valuable closures.

The differential invariants derived are used to construct invariant hyperdiffusion terms in order to model the behavior of two-dimensional freely decaying turbulence. The resulting enstrophy spectrum exhibits an arc of approximate $k^{-1}$ slope which is the theoretically derived shape for the postulated enstrophy inertial range. It should be stressed, though, that the obtained enstrophy spectrum should be taken with a pinch of salt. Since the derivation of the theoretical form of the spectra in~\cite{batc69Ay,krai67Ay} it has been tried in numerous studies to obtain these spectra in numerical simulations. Although results often vary, spectra are found with a steeper slope than the predicted $k^{-1}$ curve as described in~\cite{benz86Ay,benz88Ay,brac00Ay,legr88Ay,malt91Ay,sant89Ay}. It seems to be generally agreed today that the presence of the stable coherent vortices, which is the main feature of two-dimensional turbulence, has a strong impact on the derived enstrophy spectra. This holds in the case of turbulence both on the $\ff$-plane and on the beta-plane. The introduction of an invariant hyperdiffusion-like term certainly complicates the situation as diffusion then is coupled \emph{nonlinearly} to the vorticity equation. On the other hand, it was indicated that the presence of the beta-term in the vorticity equation allows for a nonlocal transfer of anisotropy from the larger to the smaller scales~\cite{malt91Ay}. A nonlinear diffusion term has the potential to support such a nonlocal scale interaction and thereby serves as a potential parameterization scheme for numerical models. It should be stressed in this context that in all the simulations we have carried out, the rate of energy dissipation was lower than using classical hyperdiffusion even in quite low-resolution numerical experiments.

Apart from the discussion above, the possibility of constructing hyperdiffusion-like enstrophy sink terms that lead to scale invariant enstrophy spectra seems to be a valuable property for itself. It is precisely the scale invariance of the Euler equations that is used to predict the behavior of two-dimensional turbulence in the inertial range and therefore the availability of dissipative versions of the vorticity equation having the same invariance properties as the inviscid vorticity equation might be a general advantage. Heuristically, one can expect that an invariant closure scheme should be better adapted for the problem of reproducing features that have been derived using symmetries (as the isotropic enstrophy spectrum), similarly as an invariant discretization scheme often reproduces better invariant exact solutions of a differential equation than non-invariant discretization schemes~\cite{rebe11Ay}. This assumption is supported by the proved relevance of Lie symmetries in turbulence theory~\cite{ober01Ay}. The results obtained in the present paper do not contradict this assumption, keeping in mind especially that the premises invoked to obtain the theoretical form of the spectra are at present under revision. In this context, it should again be stressed that there is a multitude of invariant parameterization schemes or invariant diffusion terms that can be coupled to the vorticity equation on the beta-plane. The fact that already the simplest invariantized version~\eqref{eq:BetaPlaneEquationInvariantHyperdiffusion} of the hyperdiffusion term (which has obvious weaknesses) shows quite good properties in the course of our numerical tests is a motivating result which is worth pointing out. Nevertheless, in order to verify and better assess the ability of invariant hyperdiffusion schemes to model turbulence on the beta-plane, further theoretical and numerical studies must be carried out.

The method we propose in this paper is fully generalizable. 
It is the number of variables of a model and its symmetry group that determine whether the method is computationally more complicated to realize, 
but this complication is not conceptual.  
Thus, the relative simplicity of constructing diffusion schemes that are invariant under the entire maximal Lie invariance group is a particular feature of the beta-plane vorticity equation, which is computationally more involved for vorticity dynamics on the \mbox{$\ff$-plane}. The complication with the latter model is that the corresponding maximal Lie invariance pseudogroup~$G_0$ is even wider than~$G_1$. This makes it much harder to derive reasonably simple closure schemes that are invariant under the entire pseudogroup~$G_0$, see the discussion in~\cite{popo10Cy}, where a generating set of differential invariants of~$G_0$ and a complete set of its independent operators of invariant differentiation are determined. A~possible remedy for this complication is to consider closure schemes that are invariant only under certain subgroups of the maximal Lie invariance pseudogroup of the $\ff$-plane equation. As highlighted in the present paper, the selection of such subgroups can be justified for physical reasons when boundaries come into play.

Another novel feature of the present paper is the explicit inclusion of conservation laws in invariant closure schemes. The chance of constructing such conservative invariant parameterization schemes is of obvious physical relevance. For physical processes that do not violate particular conservation laws, it is natural to require the associated parameterization to be also conservative. It was demonstrated in the paper for the vorticity equation on the beta-plane that the concepts of invariant and conservative parameterization schemes can be united to yield closure ansatzes that preserve both all the symmetries and certain conservation laws of this equation. The construction of further invariant conservative closure schemes as well as their exhaustive testing will be a next major challenge in the application of ideas of group analysis to the parameterization problem.

\section*{Acknowledgements}
Some of the numerical simulations were carried out using the infrastructure of the Institute of Meteorology and Geophysics of the University of Vienna.
AB acknowledges the assistance of Professor Leopold Haimberger in carrying out these simulations.
The authors are grateful to Jean-Christophe Nave, Artur Sergyeyev, Matthias Sommer, Miguel Teixeira, Pavel Winternitz and Jun-Ichi Yano for valuable discussions. We thank the two anonymous reviewers for their helpful suggestions and comments, which led to an improved version of this manuscript.
This research was supported by the Austrian Science Fund (FWF), projects J3182--N13 (AB), P20632 (ROP) and P25064 (EDSCB). The present work is carried out within the framework of COST Action ES0905.

\appendix

\section{Symmetries of the vorticity equation on the beta-plane}\label{sec:AppendixSymmetriesVorticityEquation}

We aim to detail the computation of the maximal Lie invariance algebra $\mathfrak g_1$ of the vorticity equation~\eqref{eq:VorticityEquation} here. Full expositions on finding Lie symmetries of differential equations can be found in the standard textbooks~\cite{andr98Ay,blum89Ay,olve86Ay,ovsi82Ay}. More details on the symmetries (and exact solutions) of the vorticity equation are presented in~\cite{bihl09Ay}.

Given a generator
\begin{equation}\label{eq:VectorFieldVorticityEquation}
 Q=\tau(t,x,y,\psi)\p_t+\xi(t,x,y,\psi)\p_x+\eta(t,x,y,\psi)\p_y+\varphi(t,x,y,\psi)\p_\psi.
\end{equation}
of a one-parameter point symmetry group of the vorticity equation
\[
 \Delta=\zeta_t+\psi_x\zeta_y-\psi_y\zeta_x+\beta\psi_x=0, \quad \zeta=\psi_{xx}+\psi_{yy},
\]
the infinitesimal invariance criterion~\cite{olve86Ay,ovsi82Ay} implies
$
 Q_3(\Delta)=0,
$
which has to hold on the manifold $\Delta=0$, where $Q_3$ denotes the third prolongation of the vector field $Q$.
Explicitly, the prolonged vector field $Q_3$ is defined by
$
 Q_3=Q+\sum_{0<|\alpha|\leqslant3}\varphi^\alpha\p_{\psi_\alpha}
$
and the coefficients of $Q_3$ are derived from the general prolongation formula,
\begin{equation}\label{eq:GeneralProlongationFormula}
\varphi^\alpha=\DD_t^{\alpha_1}\DD_x^{\alpha_2}\DD_y^{\alpha_3}(\varphi-\tau\psi_{\delta_1}-\xi\psi_{\delta_2}-\eta\psi_{\delta_3})+\tau\psi_{\alpha+\delta_1}+\xi\psi_{\alpha+\delta_2}+\eta\psi_{\alpha+\delta_3}.
\end{equation}
Here we use the notation introduced in the beginning of Section~\ref{sec:DifferentialInvariantsOfVorticityEquation}. Then the condition $Q_3(\Delta)=0$ expands to
\[
 \varphi^{120}+\varphi^{102}+\varphi^{010}\zeta_y+\psi_x(\varphi^{021}+\varphi^{003})- \varphi^{001}\zeta_x-\psi_y(\varphi^{030}+\varphi^{012})+\beta\varphi^{010}=0,
\]
and the constraint that $Q_3(\Delta)=0$ has to hold only on the manifold of $\Delta=0$ is taken into account by substituting $\psi_{txx}=-\psi_{tyy}-\psi_x\zeta_y+\psi_y\zeta_x-\beta\psi_x$ wherever $\psi_{txx}$ occurs.
As the coefficients of $Q$ are only functions of $t$, $x$, $y$ and $\psi$, the expanded condition can be split with respect to the various derivatives of $\psi$. This splitting yields the determining equations for the coefficients of the vector field $Q$,
\begin{align}\label{eq:DeterminingEquationsVorticityEquation}
\begin{split}
&\tau_x=\tau_y=\tau_\psi=\xi_y=\xi_\psi=\eta_t=\eta_x=\eta_\psi=\varphi_x=0,\\
&\xi_x=\eta_y=-\tau_t,\quad \varphi_y=-\xi_t,\quad \varphi_\psi=-3\tau_t.
\end{split}
\end{align}
The general solution of this system of determining equations reads
\[
 \tau=c_1t+c_2,\quad \xi=-c_1x+\tilde f(t),\quad \eta=-c_1y+c_3,\quad \varphi=-3c_1\psi-\tilde f_ty+\tilde g(t),
\]
where $\tilde f$ and $\tilde g$ run through the set of smooth functions of $t$. Thus, the maximal Lie invariance algebra of infinitesimal symmetries of the barotropic vorticity equation on the beta-plane is spanned by the vector fields
\[
\DDD=t\p_t-x\p_x-y\p_y-3\psi\p_\psi,\quad \p_t, \quad \p_y, \quad \XX(\tilde f)=\tilde f(t)\p_x-\tilde f_t(t)y\p_\psi \quad \ZZ(\tilde g)=\tilde g(t)\p_\psi.
\]

\section{Algebra of differential invariants for the vorticity equation}\label{sec:AppendixStructureAlgebraDifferentialInvariants}

In this appendix we present the details for the proof of Theorem~\ref{thm:BasisOfDifferentialInvariantsBetaPlaneEquation} which exhaustively describes the algebra of differential invariants for the maximal Lie invariance pseudogroup of the barotropic vorticity equation on the beta-plane.

A complete set of independent operators of invariant differentiation is derived by invariantization of the usual operators of total differentiation, yielding
\begin{equation}\label{eq:InvariantDifferentiationOperatorsBetaPlaneEquation}
 \DD^{\rm i}_t=\frac{1}{\sqrt{|\psi_x|}}(\DD_t-\psi_y\DD_x),\quad \DD^{\rm i}_x=\sqrt{|\psi_x|}\DD_x,\quad \DD^{\rm i}_y=\sqrt{|\psi_x|}\DD_y.
\end{equation}
This is practically realized via substituting the expressions~\eqref{eq:MovingFrameBetaPlaneEquation} for the pseudogroup parameters into
the implicit differentiation operators~\eqref{eq:ImplicitDifferentiationOperatorsBetaPlaneEquation}.
Any operator of invariant differentiation related to the pseudogroup~$G_1$ is locally a combination of the operators~\eqref{eq:InvariantDifferentiationOperatorsBetaPlaneEquation}
with functional coefficients depending only on differential invariants of~$G_1$.
The commutation relations between the operators $\DD^{\rm i}_t$, $\DD^{\rm i}_x$ and $\DD^{\rm i}_y$ are
\begin{gather}\label{eq:CommutationRelsBetweenOpsOfInvDiffForBVE}
\begin{split}
&[\DD^{\rm i}_t,\DD^{\rm i}_x]=\frac\varepsilon2I_{020}\DD^{\rm i}_t+\left(I_{011}+\frac\varepsilon2I_{110}\right)\DD^{\rm i}_x,\\
&[\DD^{\rm i}_t,\DD^{\rm i}_y]=\frac\varepsilon2I_{011}\DD^{\rm i}_t+I_{002}\DD^{\rm i}_x+\frac\varepsilon2I_{110}\DD^{\rm i}_y,\\
&[\DD^{\rm i}_x,\DD^{\rm i}_y]=\frac\varepsilon2I_{020}\DD^{\rm i}_y-\frac\varepsilon2I_{011}\DD^{\rm i}_x.
\end{split}
\end{gather}

In order to completely describe the algebra of differential invariants of~$G_1$, it remains to establish a basis of differential invariants
such that any differential invariant of~$G_1$ can be represented as a function of basis elements and their invariant derivatives.
It is also necessary to compute a complete system of syzygies between basis invariants.
For this aim, we will evaluate the recurrence relations between the normalized differential invariants and the differentiated differential invariants as detailed in~\cite{cheh08Ay,olve08Ay}.
The starting point for the application of the general algorithm to the maximal Lie invariance pseudogroup~$G_1$ of the vorticity equation on the beta-plane is the system of determining equations for the coefficients of a vector field~\eqref{eq:VectorFieldVorticityEquation} from the maximal Lie invariance algebra of Eq.~\eqref{eq:VorticityEquation}, which is given through system~\eqref{eq:DeterminingEquationsVorticityEquation}. Consider the prolonged operator $Q_\infty=Q+\sum_{|\alpha|>0}\varphi^\alpha\p_{\psi_\alpha}$.
The coefficients of~$Q_\infty$ are calculated by the standard prolongation formula~\eqref{eq:GeneralProlongationFormula}. In view of the determining equations, the coefficients $\varphi^\alpha$ take the form
\[
\varphi^\alpha=(\alpha_2+\alpha_3-\alpha_1-3)\tau_t\psi_\alpha
-\sum_{k=1}^{\alpha_1}\binom{\alpha_1}{k}\xi_{(k)}\psi_{\alpha-k\delta_1+\delta_2}+
\left\{\begin{array}{ll}-\xi_{(\alpha_1+1)},&\alpha_2=0,\ \alpha_3=1\\ \varphi_{(\alpha_1)},&\alpha_2=\alpha_3=0\end{array}\right\},
\]
\looseness=-1
where $\xi_{(k)}=\p^k\xi/\p t^k$ and $\varphi_{(k)}=\p^k\varphi/\p t^k$, $k=0,1,2,\dots$.
We collect the coefficients of~$Q$ and their derivatives appearing in the expressions for the prolonged coefficients of~$Q$
and denote the associated invariantized objects, which are differential forms, as
$\hat\tau^0=\iota(\tau)$, $\hat\tau^1=\iota(\tau_t)$, $\hat\xi^k=\iota(\xi_{(k)})$, $\hat\eta=\iota(\eta)$ and $\hat\varphi^k=\iota(\varphi_{(k)})$.
In the course of the normalization~\eqref{eq:NormalizationBetaPlaneEquation} the invariantized counterparts $\hat\varphi^\alpha=\iota(\varphi^\alpha)$ of the prolonged coefficients of~$Q$ are
\begin{gather*}
\hat\varphi^{j00}=\hat\varphi^j-\varepsilon\hat\xi^j-\sum_{k=1}^{j-1}\binom{j}{k}I_{j-k,10}\hat\xi^k \quad\mbox{if}\quad j>0,\quad
\hat\varphi^{j01}=-\hat\xi^{j+1}-\sum_{k=1}^j\binom{j}{k}I_{j-k,11}\hat\xi^k,\\
\hat\varphi^\alpha=(\alpha_2+\alpha_3-\alpha_1-3)I_\alpha\hat\tau^1-\sum_{k=1}^{\alpha_1}\binom{\alpha_1}{k}I_{\alpha-k\delta_1+\delta_2}\hat\xi^k  \quad\mbox{if}\quad \alpha_2>0 \quad\mbox{or}\quad \alpha_3>1.
\end{gather*}
For lower values of~$|\alpha|$, $0<|\alpha|\leqslant3$, we calculate
\begin{gather*}
\hat\varphi^{100}= \hat\varphi^1 -\varepsilon\hat\xi^1,\quad
\hat\varphi^{010}=-2\hat\tau^1,\quad
\hat\varphi^{001}=-\hat\xi^1,\\[.5ex]
\hat\varphi^{200}=\hat\varphi^2-\varepsilon\hat\xi^2-2I_{110}\hat\xi^1,\quad
\hat\varphi^{110}=-3I_{110}\hat\tau^1-I_{020}\hat\xi^1,\quad
\hat\varphi^{101}=-\hat\xi^2-I_{011}\hat\xi^1,\\[.5ex]
\hat\varphi^{020}=-I_{020}\hat\tau^1,\quad
\hat\varphi^{011}=-I_{011}\hat\tau^1,\quad
\hat\varphi^{002}=-I_{002}\hat\tau^1,\\[.5ex]
\hat\varphi^{300}=\hat\varphi^3-\varepsilon\hat\xi^3-3I_{110}\hat\xi^2-3I_{210}\hat\xi^1,\\[.5ex]
\hat\varphi^{210}=-4I_{210}\hat\tau^1-I_{020}\hat\xi^2-2I_{120}\hat\xi^1,\quad
\hat\varphi^{201}=-\hat\xi^3-I_{011}\hat\xi^2-2I_{111}\hat\xi^1,\\[.5ex]
\hat\varphi^{120}=-2I_{120}\hat\tau^1- I_{030}\hat\xi^1,\quad
\hat\varphi^{111}=-2I_{111}\hat\tau^1- I_{021}\hat\xi^1,\quad
\hat\varphi^{102}=-2I_{102}\hat\tau^1- I_{012}\hat\xi^1,\\[.5ex]
\hat\varphi^{030}=\hat\varphi^{021}=\hat\varphi^{012}=\hat\varphi^{003}=0.
\end{gather*}
From the recurrence relations for the phantom invariants $H^0=\iota(t)$, $H^1=\iota(x)$, $H^2=\iota(y)$, $I_{i00}=\iota(\psi_{i00})$, $I_{i01}=\iota(\psi_{i01})$, $i=0,1,\dots$, and $I_{010}=\iota(\psi_{010})$, which are
\begin{gather*}
\dddh H^0=\omega^1+\hat\tau^0=0,\ \
\dddh H^1=\omega^2+\hat\xi^0=0,\ \
\dddh H^2=\omega^3+\hat\eta=0,\ \
\dddh I_{000}=\omega^2+\hat\varphi^0=0,\\[.5ex]
\dddh I_{j00}=I_{j10}\omega^2+\hat\varphi^j-\varepsilon\hat\xi^j-\sum_{k=1}^{j-1}\binom{j}{k}I_{j-k,10}\hat\xi^k=0, \quad j=1,2,\dots,\\
\dddh I_{j01}=I_{j11}\omega^2+I_{j02}\omega^3-\hat\xi^{j+1}-\sum_{k=1}^j\binom{j}{k}I_{j-k,11}\hat\xi^k=0,\quad j=0,1,\dots,\\[1.5ex]
\dddh I_{010}=I_{110}\omega^1+I_{020}\omega^2+I_{011}\omega^3-2\hat\tau^1=0,
\end{gather*}
where $\omega^1=\iota(\ddd t)$, $\omega^2=\iota(\ddd x)$ and $\omega^3=\iota(\ddd y)$,
we derive expressions for the invariantized Maurer--Cartan forms
\begin{gather*}
\hat\tau^0=-\omega^1,\quad
\hat\xi^0=-\omega^2,\quad
\hat\eta = -\omega^3,\quad
\hat\varphi^0=-\omega^2,\quad
\hat\tau^1=\tfrac12(I_{110}\omega^1+I_{020}\omega^2+I_{011}\omega^3),\\
\hat\xi^j=I_{j-1,11}\omega^2+I_{j-1,02}\omega^3-\sum_{k=1}^{j-1}\binom{j-1}{k}I_{j-k-1,11}\hat\xi^k,\,\\
\hat\varphi^j=-I_{j10}\omega^2+\varepsilon\hat\xi^j+\sum_{k=1}^{j-1}\binom{j}{k}I_{j-k,10}\hat\xi^k,
\end{gather*}
$j=1,2,\dots$.
The forms $\hat\xi^j$ should be calculated recursively starting from $j=1$. Thus,
\begin{gather*}
\hat\xi^1=I_{011}\omega^2+I_{002}\omega^3,\\[1ex]
\hat\xi^2=(I_{111}-I_{011}^2)\omega^2+(I_{102}-I_{011}I_{002})\omega^3,\\[1ex]
\hat\xi^3=(I_{211}-3I_{011}I_{111}+I_{111}^3)\omega^2+(I_{202}-3I_{011}I_{102}+I_{011}^2I_{002})\omega^3,\quad \dots\quad.
\end{gather*}
In general,  $\hat\xi^j=\hat\xi^{j,2}\omega^2+\hat\xi^{j,3}\omega^3$, where the coefficients $\hat\xi^{j,2}$ and $\hat\xi^{j,3}$ are expressed in terms of normalized invariants~$I_\alpha$ with $|\alpha|\leqslant j+1$.

The recurrence relations for non-phantom normalized invariants correspondingly read
\begin{align*}
\dddh  I_{\alpha_1\alpha_2\alpha_3} = {}&I_{\alpha+\delta_1}\omega^1+I_{\alpha+\delta_2}\omega^2+I_{\alpha+\delta_3}\omega^3+(\alpha_2+\alpha_3-\alpha_1-3)I_\alpha\hat\tau^1\\
&-\sum_{k=1}^{\alpha_1}\binom{\alpha_1}{k}I_{\alpha-k\delta_1+\delta_2}\hat\xi^k
\quad\mbox{if}\quad \alpha_2>0 \quad\mbox{or}\quad \alpha_3>1.
\end{align*}
As by definition $\dddh  F=(\DD_t^{\rm i}F)\omega^1+(\DD_x^{\rm i}F)\omega^2+(\DD_y^{\rm i}F)\omega^3$,
the above recurrence relations can be split into a list of equations for first-order invariant derivatives of normalized differential invariants $I_\alpha$ with $\alpha_2>0$ or $\alpha_3>1$ by taking into account the expressions for the invariantized Maurer--Cartan forms:
\begin{gather}\label{eq:DifferentiaredNormalizedInvsOfBVE}
\begin{split}
&\DD^{\rm i}_tI_\alpha=I_{\alpha+\delta_1}+\frac{\alpha_2+\alpha_3-\alpha_1-3}2I_{110}I_\alpha,\\[1ex]
&\DD^{\rm i}_xI_\alpha=I_{\alpha+\delta_2}+\frac{\alpha_2+\alpha_3-\alpha_1-3}2I_{020}I_\alpha
-\sum_{k=1}^{\alpha_1}\binom{\alpha_1}{k}I_{\alpha-k\delta_1+\delta_2}\hat\xi^{k,2},\\
&\DD^{\rm i}_yI_\alpha=I_{\alpha+\delta_3}+\frac{\alpha_2+\alpha_3-\alpha_1-3}2I_{011}I_\alpha
-\sum_{k=1}^{\alpha_1}\binom{\alpha_1}{k}I_{\alpha-k\delta_1+\delta_2}\hat\xi^{k,3}.
\end{split}
\end{gather}
We only present the closed expressions for the first-order invariant derivatives of~$I_\alpha$ with $|\alpha|\leqslant3$:
\begin{gather*}
\DD^{\rm i}_tI_{110}=I_{210}-\tfrac32I_{110}^2,                    \quad
\DD^{\rm i}_xI_{110}=I_{120}-\tfrac32I_{110}I_{020}-I_{011}I_{020},\\[-.1ex]
\DD^{\rm i}_yI_{110}=I_{111}-\tfrac32I_{110}I_{011}-I_{020}I_{002},\\[-.1ex]
\DD^{\rm i}_tI_{020}=I_{120}-\tfrac12I_{110}I_{020}, \quad
\DD^{\rm i}_xI_{020}=I_{030}-\tfrac12I_{020}^2,      \quad
\DD^{\rm i}_yI_{020}=I_{021}-\tfrac12I_{011}I_{020}, \\[-.1ex]
\DD^{\rm i}_tI_{011}=I_{111}-\tfrac12I_{110}I_{011}, \quad
\DD^{\rm i}_xI_{011}=I_{021}-\tfrac12I_{011}I_{020}, \quad
\DD^{\rm i}_yI_{011}=I_{012}-\tfrac12I_{011}^2,      \\[-.1ex]
\DD^{\rm i}_tI_{002}=I_{102}-\tfrac12I_{110}I_{002}, \quad
\DD^{\rm i}_xI_{002}=I_{012}-\tfrac12I_{020}I_{002}, \quad
\DD^{\rm i}_yI_{002}=I_{003}-\tfrac12I_{011}I_{002}, \\[-.1ex]
\DD^{\rm i}_tI_{210}=I_{310}-2I_{110}I_{210},                                                 \quad
\DD^{\rm i}_xI_{210}=I_{220}-2I_{020}I_{210}-2I_{011}I_{120}+(I_{011}^2-I_{111})I_{020},      \\[-.1ex]
\DD^{\rm i}_yI_{210}=I_{211}-2I_{011}I_{210}-2I_{002}I_{120}+(I_{002}I_{011}-I_{102})I_{020}, \\[-.1ex]
\DD^{\rm i}_tI_{201}=I_{301}-2I_{110}I_{201},                                                 \quad
\DD^{\rm i}_xI_{201}=I_{211}-2I_{020}I_{201}-3I_{001}I_{111}+I_{011}^3,                       \\[-.1ex]
\DD^{\rm i}_yI_{201}=I_{202}-2I_{011}I_{201}-2I_{002}I_{111}-I_{011}I_{102}+I_{002}I_{011}^2, \\[-.1ex]
\DD^{\rm i}_tI_{120}=I_{220}-I_{110}I_{120},                \quad
\DD^{\rm i}_xI_{120}=I_{130}-I_{020}I_{120}-I_{011}I_{030}, \\[-.1ex]
\DD^{\rm i}_yI_{120}=I_{121}-I_{011}I_{120}-I_{002}I_{030}, \\[-.1ex]
\DD^{\rm i}_tI_{111}=I_{211}-I_{110}I_{111},                \quad
\DD^{\rm i}_xI_{111}=I_{121}-I_{020}I_{111}-I_{011}I_{021}, \\[-.1ex]
\DD^{\rm i}_yI_{111}=I_{112}-I_{011}I_{111}-I_{002}I_{021}, \\[-.1ex]
\DD^{\rm i}_tI_{102}=I_{202}-I_{110}I_{102},                \quad
\DD^{\rm i}_xI_{102}=I_{112}-I_{020}I_{102}-I_{011}I_{012}, \\[-.1ex]
\DD^{\rm i}_yI_{102}=I_{103}-I_{011}I_{102}-I_{002}I_{012}, \\[-.1ex]
\DD^{\rm i}_tI_{030}=I_{130}, \quad
\DD^{\rm i}_xI_{030}=I_{040}, \quad
\DD^{\rm i}_yI_{030}=I_{031}, \\[-.1ex]
\DD^{\rm i}_tI_{021}=I_{121}, \quad
\DD^{\rm i}_xI_{021}=I_{031}, \quad
\DD^{\rm i}_yI_{021}=I_{022}, \\[-.1ex]
\DD^{\rm i}_tI_{012}=I_{112}, \quad
\DD^{\rm i}_xI_{012}=I_{022}, \quad
\DD^{\rm i}_yI_{012}=I_{013}, \\[-.1ex]
\DD^{\rm i}_tI_{003}=I_{103}, \quad
\DD^{\rm i}_xI_{003}=I_{013}, \quad
\DD^{\rm i}_yI_{003}=I_{004}.
\end{gather*}
In principle, it is possible to read off the generating differential invariants from the above split recurrence relations.
The expressions for $I_{\alpha+\delta_1}$, $I_{\alpha+\delta_2}$ and $I_{\alpha+\delta_3}$ derived from~\eqref{eq:DifferentiaredNormalizedInvsOfBVE}
only involve first-order invariant derivatives of~$I_\alpha$ and normalized invariants of orders not greater than $|\alpha|$.
This implies that a generating set of differential invariants consists of invariantized derivatives which are minimal with respect to the usual partial ordering of derivatives and
are not phantom invariants. We have four such minimal elements,
\[
  I_{110}=\frac{\psi_{tx}-\psi_y\psi_{xx}}{\sqrt{|\psi_x|^3}},\quad
  I_{020}=\frac{\psi_{xx}}{\sqrt{|\psi_x|}},\quad
  I_{011}=\frac{\psi_{xy}}{\sqrt{|\psi_x|}},\quad
  I_{002}=\frac{\psi_{yy}}{\sqrt{|\psi_x|}}.
\]
All the other invariantized derivatives are expressed via invariant derivatives of~$I_{110}$, $I_{020}$, $I_{011}$ and $I_{002}$.
As was indicated above, not all differentiated differential invariants are necessarily functionally independent. This fact is encoded in syzygies of the algebra of differential invariants. Taking into account these syzygies can further reduce the number of generating differential invariants thereby allowing one a more concise description of the basis of differential invariants.
In the present case, we find the following lower-order syzygies:
\begin{gather*}
\DD^{\rm i}_tI_{011} - \DD^{\rm i}_yI_{110} = I_{110}I_{011}+I_{020}I_{002},\\
\DD^{\rm i}_tI_{020} - \DD^{\rm i}_xI_{110} = I_{020}(I_{110}+I_{011}),\\
\DD^{\rm i}_yI_{011} - \DD^{\rm i}_xI_{002} = \tfrac12I_{020}I_{002}-\tfrac12I_{011}^2,\\
\DD^{\rm i}_xI_{011} - \DD^{\rm i}_yI_{020} = 0,
\\
(\DD^{\rm i}_y)^2I_{110} - \DD^{\rm i}_t\DD^{\rm i}_xI_{002} =
\tfrac12(\DD^{\rm i}_t-I_{011})(I_{020}I_{002})-(\DD^{\rm i}_y+I_{011})(\tfrac32I_{110}I_{011}+I_{020}I_{002})\\\qquad
-I_{011}\DD^{\rm i}_yI_{110}-I_{002}\DD^{\rm i}_yI_{020},
\\
(\DD^{\rm i}_y)^2I_{020} - (\DD^{\rm i}_x)^2I_{002} = \tfrac12\DD^{\rm i}_x(I_{020}I_{002})-\tfrac12\DD^{\rm i}_y(I_{011}I_{020}).
\end{gather*}
From the two first syzygies we can express the invariants $I_{011}$ and $I_{002}$ via invariant derivatives of $I_{110}$ and $I_{020}$,
\begin{gather*}
I_{011}=\frac{\DD^{\rm i}_tI_{020}-\DD^{\rm i}_xI_{110}}{I_{020}}-I_{110},\\
I_{002}=\frac1{I_{020}}(\DD^{\rm i}_t-I_{110})\left(\frac{\DD^{\rm i}_tI_{020}-\DD^{\rm i}_xI_{110}}{I_{020}}-I_{110}\right)-\frac{\DD^{\rm i}_yI_{110}}{I_{020}}.
\end{gather*}

Another way of finding relations between generating invariants is to use the commutation relations between the operators of invariant differentiation.
Evaluating each equality from~\eqref{eq:CommutationRelsBetweenOpsOfInvDiffForBVE} on an element~$I$ from the above generating set,
we obtain a system of linear algebraic equations with respect to the other elements of these sets,
which can be solved on the domain of the jet space where the determinant of the matrix associated with the system does not vanish.
It is convenient to choose, e.g., $I=I_{020}$. Then, we derive the representations
\begin{gather*}
I_{011}=\frac{
I_{020}\DD^{\rm i}_yI_{020}-2\varepsilon[\DD^{\rm i}_x,\DD^{\rm i}_y]I_{020}}
{\DD^{\rm i}_xI_{020}},
\\
I_{110}=\frac{
2\varepsilon[\DD^{\rm i}_t,\DD^{\rm i}_x]I_{020}-I_{020}\DD^{\rm i}_tI_{020}
}{\DD^{\rm i}_xI_{020}}
-2\varepsilon I_{011},
\\
I_{002}=
\frac{[\DD^{\rm i}_t,\DD^{\rm i}_y]I_{020}}{\DD^{\rm i}_xI_{020}}
-\frac\varepsilon2\frac{\DD^{\rm i}_tI_{020}}{\DD^{\rm i}_xI_{020}}I_{011}
-\frac\varepsilon2\frac{\DD^{\rm i}_yI_{020}}{\DD^{\rm i}_xI_{020}}I_{110},
\end{gather*}
which are defined on the domain~$\Omega_1$ of the jet space where $\DD^{\rm i}_xI_{020}\ne0$, i.e., $\DD_x^{\,2}(\sqrt{|\psi_x|}\,)\ne0$.

As a result, it is straightforward to establish Theorem~\ref{thm:BasisOfDifferentialInvariantsBetaPlaneEquation}.


\end{document}